\documentclass[twocolumn,superscriptaddress,amssymb, nobibnotes, aps, prb]{revtex4-2}
\usepackage{amsmath}
\usepackage{graphicx}
\usepackage{amsfonts}
\usepackage{amssymb}
\usepackage{bm}
\usepackage{dsfont}
\usepackage{array}
\usepackage{subfig}
\usepackage{epsfig,float,afterpage}
\usepackage[colorlinks,linkcolor=blue,citecolor=blue,urlcolor=blue]{hyperref}
\usepackage[usenames,dvipsnames]{color}

\newcommand{\beq}{\begin{equation}}
\newcommand{\eeq}{\end{equation}}
\newcommand{\bes}{\begin{subequations}}
\newcommand{\ees}{\end{subequations}}
\newcommand{\bea}{\begin{eqnarray}}
\newcommand{\eea}{\end{eqnarray}}
\newcommand{\ba}{\begin{array}}
\newcommand{\ea}{\end{array}}
\newcommand{\beqn}{\begin{eqnarray*}}
\newcommand{\eeqn}{\end{eqnarray*}}

\newcommand{\ra}{\rangle}

\newcommand{\dg}{\dagger}

\newcommand{\Ha}{\hat{a}}
\newcommand{\Had}{\hat{a}^\dag}

\newcommand{\Hs}{\hat{\sigma}}
\newcommand{\Hsd}{\hat{\sigma}^\dag}

\newcommand{\HN}{\hat{N}}
\newcommand{\Hh}{\hat{H}}
\newcommand{\MM}{\mathcal{M}}
\newcommand{\MN}{\mathcal{N}}
\newcommand{\tr}{\text{tr}}

\def\nn{\nonumber}

\begin{document}
\title{Many-body quantum chaos in mixtures of multiple species}
\author{Vijay Kumar and Dibyendu Roy}
\affiliation{Raman Research Institute, Bangalore 560080, India}
\begin{abstract}
 We study spectral correlations in many-body quantum mixtures of fermions, bosons, and qubits with periodically kicked spreading and mixing of species. We take two types of mixing, namely, Jaynes-Cummings and Rabi, respectively, satisfying and breaking the conservation of a total number of species. We analytically derive the generating Hamiltonians whose spectral properties determine the spectral form factor in the leading order. We further analyze the system-size $(L)$ scaling of Thouless time $t^*$, beyond which the spectral form factor follows the prediction of random matrix theory. The $L$-dependence of $t^*$ crosses over from $\log L$ to $L^2$ with an increasing Jaynes-Cummings mixing between qubits and fermions or bosons in a finite-sized chain, and it finally settles to $t^* \propto \mathcal{O}(L^2)$ in the thermodynamic limit for any mixing strength. The Rabi mixing between qubits and fermions leads to $t^*\propto \mathcal{O}(\log L)$, previously predicted for single species of qubits or fermions without total number conservation. 

\end{abstract}

\maketitle

A series of recent microscopic studies has explored quantum chaos and spectral correlations in periodically driven (Floquet) many-body systems \cite{KosPRX2018,BertiniPRL2018,ChanPRX2018,ChanPRL2018,FriedmanPRL2019,Bertini2019PRL,RoyPRE2020,Bertini2021CMP,Garratt2021PRX,LiPRL2021,KosPRL2021,Moudgalya21,RoyPRE2022,Swingle22,Winer22,Liao22,Liao2022,Dag23} to show the emergence of universal random matrix theory (RMT) description of the spectral form factor (SFF) in these models by going beyond the semiclassical periodic-orbit approaches \cite{Haake2001,Berry1985}. These investigations have further strengthened our understanding of the {\it quantum chaos conjecture} \cite{BohigasPRL1984,McDonaldPRL1979,Casati1980,Berry1977,Berry1981,Sieber2001,Sieber2002,MullerPRL2004,MullerPRE2005} for describing the spectral fluctuations of many-body nonintegrable quantum systems by RMT. Till now, such microscopic derivation of SFF in many-body quantum models has been restricted to systems with single components/species, e.g., fermions, bosons, and qubits. Nature, however, is full of systems consisting of multiple species, such as the crystalline solids of electrons and phonons and the black-body radiation comprising thermal electromagnetic radiation within or surrounding a matter in thermodynamic equilibrium. Inspired by these examples, we derive the leading order contributions to SFF in various mixed many-body quantum systems with two different types of species, e.g., qubits and bosons or fermions \cite{CarlosPRL2019}.

We consider many-body quantum mixtures where a base Hamiltonian with the entries diagonal in the Fock space basis of two different species is kicked periodically by another Hamiltonian with terms consisting of mixing between two species and nearest-neighbor hopping of any one species. The diagonal entries in the base Hamiltonian include random chemical potentials and transition frequencies along with pairwise long-range interactions of one species. We consider two forms of the mixing Hamiltonian: (a) Jaynes-Cummings $(JC)$ \cite{Jaynes1963, scully1997quantum, Hartmann_2006, Angelakis2007, RoyRMP2017} and (b) Rabi $(R)$ \cite{Rabi1937,Xie2017} interaction between different species. While the $JC$ preserves the total number of excitations of both species, the $R$ does not. Thus, we have $U(1)$) symmetry in the $JC$ mixing system, which is absent for the $R$ mixing. Our models' two different components are either qubit and spinless boson or qubit and spinless fermion. Since spinless fermions are related to spin-1/2s or qubits, our results here are valid for many different types of mixture, e.g., the results for a compound model of qubits and spinless bosons are also helpful for a mix between spinless fermions and bosons. Similarly, the results for a mixture of qubits and spinless fermions apply to a mixture of spin-1/2s of different species, e.g., electrons and atomic nuclei in solids.

First, we rewrite the spectral form factor of the compound systems in terms of a bi-stochastic many-body process \cite{RoyPRE2020,RoyPRE2022} generated by an effective Hamiltonian. The effective Hamiltonian describes the leading order contributions of SFF within the random phase approximation (RPA) in the Trotter regime of small perturbation parameters. We identify symmetries of the effective Hamiltonian controlling dynamical processes for the emergence of RMT behavior in these models \cite{RoyPRE2022, Agarwal2023}. These symmetries are important in determining system-size $(L)$ scaling of the Thouless timescales $t^*$ beyond which the SFF has a universal RMT/COE form for our time-reversal invariant models of a circular orthogonal ensemble (COE). For $JC$ mixing, we find, $t^* \propto L^2$ when $L \to \infty$, which is a characteristics of $U(1)$-symmetric model \cite{Gharibyan2018,FriedmanPRL2019,RoyPRE2020, RoyPRE2022}. However, we show an exciting competition between the hopping and mixing of the driving Hamiltonian, leading to a crossover behavior in the $L$-dependence of $t^*$ when a finite-size system is considered. For a finite system, $t^* \propto \log L$ when the mixing strength is smaller than the hopping, and $t^* \propto L^2$ for a higher mixing strength compared to hopping. The above crossover in $L$ scaling of $t^*$ is inevitable in many experimental studies with highly controlled laboratory settings of finite size \cite{RoyRMP2017, islam2015measuring, kaufman2016quantum, bernien2017probing, Cronenberger2019, Swar2021, Joshi2022, Dag23}. For $R$ mixing between fermions and qubits, $t^* \propto \log L$ or $L^0$ for large $L$, which is similar to the single species of fermion or spin-1/2 models in the absence of $U(1)$ symmetry. In contrast to fermions or spin-1/2s, the only boson model lacking $U(1)$ symmetry shows an algebraic $L$-dependence of $t^*$ \cite{RoyPRE2022}. We offer numerical evidence that the $L$-dependence of $t^*$ for $R$ mixing between bosons and qubits seems to behave similarly to $R$ mixing between fermions and qubits.

The base (kicked) Hamiltonian $\hat{H}_0$ of our systems denotes a one-dimensional lattice of length $L$ consisting of spinless fermions or bosons and qubits with no coupling between these two entities/species.   
\bea
\hat{H}_0&=& \sum_{i=1}^L(\omega_i\hat{n}_i + \Omega_i\hat{\sigma}^{\dg}_i\hat{\sigma}_i)+\sum_{i<j} U_{ij}\hat{n}_i\hat{n}_j,
\eea
where $\hat{n}_i=\hat{a}_{i}^{\dg}\hat{a}_{i}$ is the number operator with $\hat{a}_{i}^{\dg}$ being a fermion or boson creation operator at site $i$. The raising  and lowering operators $\hat{\sigma}^{\dg}_j \equiv (\hat{\sigma}_j^x+i\hat{\sigma}^y_j)/2, \hat{\sigma}_j \equiv (\hat{\sigma}_j^x-i\hat{\sigma}^y_j)/2$ are for the qubit at site $j$. Here, $\omega_i$ and $\Omega_i$ are, respectively onsite energy/frequency of the fermion/boson and the transition frequency of the qubit at site $i$. We choose one or both of $\omega_i$ and $\Omega_i$ random as Gaussian {\em iid} variables of zero mean and finite standard deviation. We further take long-range interaction between fermions or bosons at sites $i$ and $j$ given by $U_{ij}=U_0/|i-j|^{\alpha}$ with an exponent in the interval $1<\alpha<2$. The form of $\hat{H}_0$ is fixed by minimal requirements for analytical calculation as well as physical relevance. Our analytical calculation requires the RPA and integration out of the parameters of $\hat{H}_0$, and both are met by the above choice of $\hat{H}_0$. The model with bosons and qubits and its close variants can physically represent light-matter interactions in real systems and engineered meta-materials \cite{scully1997quantum, RoyRMP2017, Xie2017} and electron-phonon interactions in crystalline solids.

The driving/kicking Hamiltonian consists of a term denoting the mixing between fermions/bosons and qubits locally and another term indicating nearest-neighbor hopping of fermions/bosons. The driving Hamiltonian with $JC$ and $R$ interactions are, respectively,
\bea
&&\hat{H}_{\text JC} = \sum_{i=1}^L g(\Had_i\Hs_i + \Hsd_i\Ha_i)+\sum_{i=1}^L(-J\hat{a}_i^{\dg}\hat{a}_{i+1}+{\rm h.c.}), \label{HamJC}\\
&&\hat{H}_{\text R} = \sum_{i=1}^L g(\Had_i+\Ha_i)(\Hs_i+\Hsd_i)+\sum_{i=1}^L(-J\hat{a}_i^{\dg}\hat{a}_{i+1}+{\rm h.c.}),\nn\\\label{HamR}
\eea
where $g$ and $J$ are the strength of mixing and hopping. The total excitation number operator, $\hat{N}=\sum_{i=1}^L(\hat{n}_i+\hat{\sigma}^{\dg}_i\hat{\sigma}_i)$, commutes with both $\hat{H}_0$ and $\hat{H}_{\text JC}$, but not with $\hat{H}_{\text R}$. Thus, the time-dependent total Hamiltonian, $\hat{H}(t)=\hat{H}_0+\hat{H}_{\text JC/R}\sum_{m \in \mathbb{Z}}\delta(t-m)$, commutes with $\hat{N}$ for $JC$ interaction but not for $R$ interaction showing the presence or absence of a $U(1)$ symmetry, which corresponds respectively to conservation or violation of the total excitation number in our models. We here use periodic boundary condition (PBC) in real space, i.e., $\hat{a}_i\equiv\hat{a}_{i+L}, \hat{\sigma}_i\equiv\hat{\sigma}_{i+L}$.

The SFF, $K(t)$, is defined as a time $(t)$ Fourier transformation of the two-point correlation of the spectral density of quasienergies, which are eigenvalues of the unitary one-cycle Floquet propagator $\hat{U}$ of our periodically driven systems. $K(t)$ can be written as \cite{KosPRX2018, RoyPRE2020}
\bea
K(t)=\langle (\tr \hat{U}^t)(\tr \hat{U}^{-t}) \rangle - (\MN_{\zeta}^{\beta})^2 \delta_{t,0}, \label{SFF}
\eea
where $\MN_{\zeta}^{\beta}$ is the dimension of the Hilbert space of the system with $\zeta=JC, R$ mixing for fermions $(\beta=F)$ and bosons $(\beta=B)$. 
Here, $\langle \dots \rangle$ denotes an average over the quench disorders $\{ \Omega_i\}$ and/or $\{\omega_i\}$. The one-cycle time-evolution operator $\hat{U}$ can be expressed as
\bea
\hat{U} = \hat{V}\hat{W}, \quad
	\hat{W} = e^{-i\Hh_0} \text{ and } \hat{V} = e^{-i\Hh_{\text JC/R}}.
\eea
We consider the basis states $|\underline{n\sigma}\rangle \equiv |n_1,\ldots,n_L \rangle \otimes |\sigma_1,\ldots,\sigma_L\rangle,$ where the occupation number of spinless fermion/boson and qubit at the lattice site $j$ are respectively given $n_j = 0,1~ (F)~ {\rm and}~  0,1,2,\ldots (B)$, and $\sigma_j = 0,1$. The total number of excitations  $N \equiv \langle \underline{n\sigma}|\HN|\underline{n\sigma}\rangle = \sum_{j=1}^L(n_j+\sigma_j)$ is conserved in the whole system only for $JC$ mixing.

For $JC$ mixing of fermions and qubits, we can distribute total excitations $N(<2L)$ among $2L$ states consisting of $L$ spatially localized qubit excitations and another $L$ spatially delocalized fermionic excitations.  Thus, the dimension of the Hilbert space for this system with $N$ excitations $\mathcal{N}_{\text JC}^{\text F}=(2L)!/((2L-N)!N!)$. We further have, $\mathcal{N}_{\text R}^{\text F}=\sum_{N=0,2..}^{2L}\mathcal{N}_{\rm JC}^{\rm F}=2^{2L-1}$, which is the dimension of the even sector of Hilbert space for $R$ mixing of fermions and qubits. 

For $JC$ mixing between bosons and qubits, the number of qubit excitations $M$ ($\equiv \sum_{j=1}^L \sigma_j$) can be $0 \leq M \leq \min(N,L)$. The total number of bosons there would be $N-M$. We can find the dimension $\MN_{\rm JC}^{\rm B}$ of the Hilbert space by summing over allowed $M$. Thus, we get
\begin{equation}
	\MN_{\text JC}^{\text B} = \sum_{M=0}^{\min(N,L)}
	\frac{L!}{M!(L-M)!}\frac{(N-M+L-1)!}{(N-M)!(L-1)!}.
\end{equation}
The Hilbert space dimension $\MN_{\text R}^{\text B}$ becomes infinite for $R$ mixing of bosons and qubits as $N$ is not conserved and has no upper bound. However, as discussed later, it is possible to introduce a truncation for a maximum number of total excitation $N_{max}$ in the lattice for numerical calculation. 

\begin{figure*}
\includegraphics[width=\linewidth]{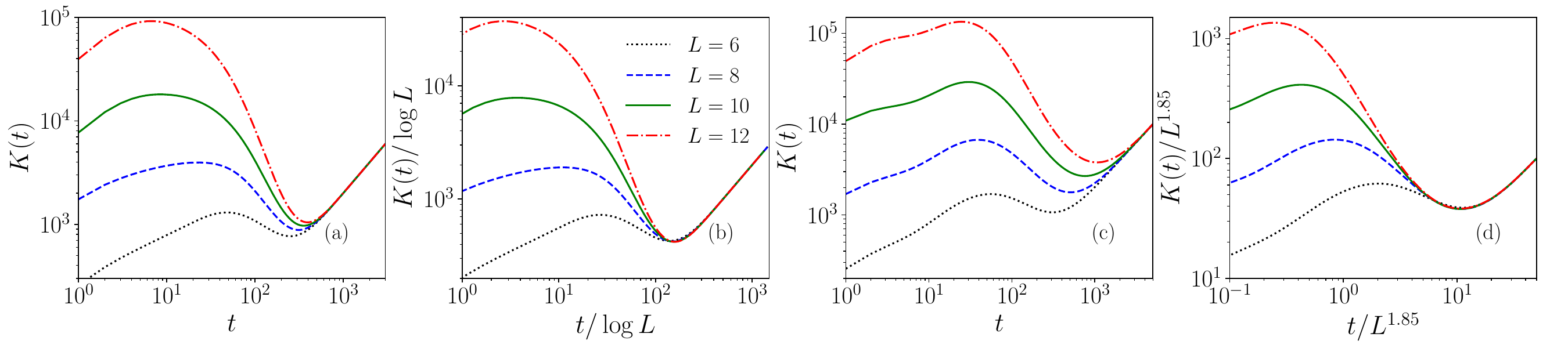}
\caption{Spectral form factor $K(t)$ using Eq.~\ref{series} for different system sizes $L$ of the kicked chain with $JC$ mixing between fermions and qubits for $g=0.1,J=0.4$ in (a,b), and $g=0.4,J=0.1$ in (c,d). We take half-filling $N/L=1/2$. In (b) and (d), we show data collapse in scaled time $t/\log L$ and $t/L^{1.85}$, respectively.}
\label{Fig1}
\end{figure*}

Both for fermionic and bosonic models, these basis states $|\underline{n\sigma}\rangle$ are eigenstates of $\Hh_0$ and $\hat{W}$, which allows us to integrate out $\Hh_0$ from $\hat{U}$ and $K(t)$ through the RPA by disorder averaging over different realizations. We further make an identity permutation approximation to achieve the following simple form  of the SFF~\cite{RoyPRE2020, RoyPRE2022} by including the leading order contributions at times $t\ll t_{\rm H} \equiv \MN_{\zeta}^{\beta}$:
$K(t) = 2t\: \tr \MM^t$, where $\MM$ is a $\mathcal{N}_{\zeta}^{\beta} \times \mathcal{N}_{\zeta}^{\beta}$ double stochastic square matrix whose elements are $\MM_{\underline{n\sigma},\underline{n\sigma}'} = |\langle \underline{n\sigma}|\hat{V}|\underline{n\sigma}'\rangle|^2$. The largest eigenvalue of $\mathcal{M}$ is one due to the unitarity of $\hat{V}$. Thus, we can write the eigenvalues of $\mathcal{M}$ as $1,\lambda_1,\lambda_2,\lambda_3,\dots$ with $1\ge |\lambda_i| \ge |\lambda_{i+1}|$. Using these eigenvalues, we express the SFF as (see Sec. I of SM\cite{SM} for a derivation):
\bea
K(t)=2t\big(1+\sum_{i=1}^{\MN_{\zeta}^{\beta}-1}\lambda_i^t\big), \label{series}
\eea
where $K(t)\simeq 2t$ is a leading order in $t/t_{\rm H}$ result of RMT/COE. The RMT/COE form of $K(t)\simeq 2t$ in a leading order appears beyond the Thouless timescales $t^*(L)$ when the contribution from the second term in Eq.~\ref{series} becomes negligible. The contribution from the second term depends on the properties of $\lambda_i$ for $i \ge 1$. We next try to understand the features of $\mathcal{M}$ and its eigenvalues. We can find Hermitian quantum Hamiltonians generating $\mathcal{M}$ in the Trotter regime of small $g,J$ for fermionic and bosonic models with $JC$ and $R$ mixing. The Hamiltonians are derived by writing $\MM$ using an element-wise commutative product (also known as the Hadamard product) of $\hat{V}$ with $\hat{V}^*$ in the basis $|\underline{n\sigma}\rangle$, and then expanding $\hat{V}$ in the Trotter regime of small parameters of $\Hh_{\rm JC}~(\Hh_{\rm R})$ up to second order in $\Hh_{\rm JC}~(\Hh_{\rm R})$. The emergent symmetries of these generating Hamiltonians control the dynamical properties of these models, such as $t^*(L)$, and they can be significantly different from the symmetries of $\Hh(t)$.

We first analyze $\mathcal{M}$ for the fermionic model with $JC$ mixing. The generating Hamiltonian for PBC is (see Sec. II of SM\cite{SM} for a derivation)
\bea
\mathcal{M}_{\text JC}^{\text F}&=&\big(1-\frac{(g^2+J^2)L}{2}\big)\mathds{1}_{\mathcal{N}_{\text JC}^{\text F}}+\sum_{i=1}^L \sum_{\nu}\Big(\frac{J^2}{2}\hat{\tau}_i^{\nu}\hat{\tau}_{i+1}^{\nu}\nn\\&+&\frac{g^2}{2}\hat{\tau}_i^{\nu}\hat{\sigma}_{i}^{\nu}\Big)+\mathcal{O}(J^4,g^4), \label{mapHamJCF}
\eea
where $\hat{\tau}^{\nu}_i$ and $\hat{\sigma}^{\nu}_i$ are the $\nu$th component of Pauli matrix at site $i$ and $\nu \in \{x,y,z\}$. Here, $\hat{\tau}^{\nu}_i$ and $\hat{\sigma}^{\nu}_i$ represent, respectively, the spinless fermions and qubits. The largest eigenvalue one of $\mathcal{M}_{\text JC}^{\text F}$ corresponds to a state in which all $\tau$ and $\sigma$ spins are polarized in one particular direction, say along $z$ axis. $\mathcal{M}_{\text JC}^{\text F}$ commutes with the operators, $\sum_{j=1}^L(\hat{\tau}^{\nu}_j+\hat{\sigma}^{\nu}_j)/2$ for $\nu \in \{x,y,z\}$, which satisfy $SU(2)$ algebra. Thus, $\mathcal{M}_{\text JC}^{\text F}$ has $SU(2)$ symmetry, which implies that there would be degenerate symmetry multiplets of the subleading eigenvalues of $\mathcal{M}_{\text JC}^{\text F}$ for different $N~(=1,2,3,\dots2L-1)$. Nevertheless, other energy eigenvalues can also appear between different descendent states for higher $N$. Since we are interested in $L$-dependence of $t^*$ at finite filling fractions $(N/L)$, the ordering of descendant states in the full spectrum of $\mathcal{M}_{\text JC}^{\text F}$ for $N>1$ is important. It can be shown that the value of $\lambda_1$ is the same for all $N$, including $N=1$ at any value of $g, J$. The eigenvalues of $\mathcal{M}_{\text JC}^{\text F}$ excluding the largest eigenvalue one for $N=1$ are
\bea
\lambda_i=1-g^2-J^2\big(1-\cos\frac{2i\pi}{L}\big)+\sqrt{J^4\big(1-\cos\frac{2i\pi}{L}\big)^2+g^4},\nn\\\label{ev1}
\eea
for $i=1,2,3\dots,L-1$. In the thermodynamic limit of $L \to \infty$, we find from Eq.~\ref{ev1}, $\lambda_1 \approx 1-(2\pi^2J^2)/L^2$ for any value of $g,J$. However, such approximation of Eq.~\ref{ev1} is also applicable for large $L~(>{\it l}_c\equiv \pi/\sin^{-1}(g/\sqrt{2}J)$, a critical length-scale depending on $g,J$) when $(1-\cos\frac{2\pi}{L}\big) \ll (g/J)^2$. We then further approximate $K(t)$ at long time $t$, $1\ll t\ll \mathcal{N}_{\text JC}^{\text F}$, by keeping up to the second largest eigenvalue $\lambda_1$ of $\mathcal{M}_{\text JC}^{\text F}$. Thus, we get for SFF
\bea
K(t)\simeq 2t(1+\lambda_1^t)\simeq 2t(1+e^{-t/t^*(L)}),\label{SFFl}
\eea
where we take the scaling of $\lambda_1$  with system size $L$ as $1-1/t^*$ and $t^*=L^2/(2\pi^2J^2)$ \cite{RoyPRE2020,RoyPRE2022}. The above $L$-dependence of $t^*$ is similar to our earlier result in Ref.~\cite{RoyPRE2020} for a $U(1)$ symmetric fermionic model without the qubits. 

However, there is another interesting parameter regime when $(1-\cos\frac{2i\pi}{L}\big) \gg (g/J)^2$ for a finite $L$, which is the case in many proposed highly controlled laboratory settings to test predictions for SFF \cite{Roy2015, Cronenberger2019, Swar2021, Joshi2022, Dag23}. For $(1-\cos\frac{2i\pi}{L}\big) \gg (g/J)^2$ at finite $L~(<{\it l}_c)$, 
 we have $\lambda_i \approx 1-g^2+(g^2/2J)^2\csc^2(i\pi/L)$ for $i=1,2,3\dots,L-1$. Therefore, the second largest eigenvalues for small $g/J$ are approximately $L-1$ fold degenerate. These features of $\lambda_i$ give a different form of SFF, and the system-size scaling of $t^*$: $K(t)\simeq 2t(1+\sum_{i=1}^{L-1}\lambda_i^t)$, which leads to $t^*\approx \mathcal{O}(\log L)$ (check Sec. II of SM\cite{SM}). Such logarithmic system size dependence of $t^*$ has been previously reported for $U(1)$ symmetry-broken models in the absence of total particle number conservation \cite{KosPRX2018}. We can also increase the value of $g/J$ at a fixed finite $L$ to access the other condition,  $(1-\cos\frac{2i\pi}{L}\big) \ll (g/J)^2$, to get the SFF in Eq.~\ref{SFFl}, and $t^* \propto L^2$. Thus, we find a crossover in system size scaling of $t^*$ with a varying scaled mixing strength $g/J$ at finite lengths in our model with $JC$ mixing between fermions and qubits.

To demonstrate two different $L$ scaling of $t^*$, we plot $K(t)$ with $t$ using Eq.~\ref{series}, which is obtained by applying the RPA and identity permutation for leading order contributions. In Figs.~\ref{Fig1}(a,c), we show $K(t)$ with $t$ for $g=0.1,J=0.4~({\it l}_c \sim 17)$ and $g=0.4,J=0.1~({\it l}_c \sim 0)$, respectively. We take the half-filled case with $N/L=1/2$. We can understand the $L$ dependence of $t^*$ for these two parameter sets by scaling $t$ and $K(t)$ by predicted $L$ dependence. For this, we plot $K(t)/\log L$ against $t/\log L$ in Fig.~\ref{Fig1}(b) and $K(t)/L^{1.85}$ against $t/L^{1.85}$ in Fig.~\ref{Fig1}(d). Figs.~\ref{Fig1}(b,d) display a nice data collapse for different $L$ at a time above $t^*$ for the universal RMT behavior of the SFF. Such data collapse confirms our above-predicted crossover of the $L$ dependence of $t^*$ with an increasing $g/J$. We could not get $t^*$ growing exactly as $L^2$ for a large $g/J$ in our numerics with limited $L$. Still, our obtained exponent $(\sim 1.85)$ in this region is close to the predicted value of $2$.  

The generating Hamiltonian for $JC$ mixing between bosons and qubits in the Trotter regime reads as (see Sec. III of SM\cite{SM} for a derivation)
\bea
\mathcal{M}_{\text JC}^{\text B}&=&\big(1+\frac{(g^2+J^2)L}{2}\big)\mathds{1}_{\mathcal{N}_{\text JC}^{\text B}}+\sum_{i=1}^L\Big(2J^2\big(\hat{K}^1_i\hat{K}^1_{i+1}\nn\\&+&\hat{K}^2_{i}\hat{K}^2_{i+1}-\hat{K}^0_i\hat{K}^0_{i+1}\big)+g^2(\hat{K}^1_i\hat{\sigma}^x_i-\hat{K}^2_{i}\hat{\sigma}^{y}_i\nn\\&+&\hat{K}^0_i-\hat{\sigma}^{\dagger}_i\hat{\sigma}_i)\Big)+\mathcal{O}(J^4,g^4), \label{mapHamJCB}
\eea
where $\hat{K}^1_j=(\hat{K}^+_j+\hat{K}^-_j)/2, \hat{K}^2_j=(\hat{K}^+_j-\hat{K}^-_j)/2i$. We define a set of local operators $\hat{K}^0_j=-(\hat{n}_j+1/2),~ \hat{K}^+_j=\hat{a}_j\sqrt{\hat{n}_j},~ \hat{K}^-_j=\sqrt{\hat{n}_j}\hat{a}^{\dagger}_j$, which satisfy the commutation relations of $SU(1,1)$ algebra at the same site, and commute otherwise: $[\hat{K}^+_i,\hat{K}^-_j]=-2\hat{K}^0_i \delta_{ij},~ [\hat{K}^0_i,\hat{K}^{\pm}_j]=\pm\hat{K}^{\pm}_i \delta_{ij}$. However, $\mathcal{M}_{\text JC}^{\text B}$ in Eq.~\ref{mapHamJCB} does not commute with $\hat{K}^\alpha = \sum_{i=1}^L \hat{K}^\alpha_i$, $\alpha\in\{+,-,0\}$ for a non-zero $g$. Thus, $\mathcal{M}_{\text JC}^{\text B}$ does not possess $SU(1,1)$ symmetry unlike the only boson model investigated in \textcite{RoyPRE2022}. Nevertheless, we find $[\mathcal{M}_{\text JC}^{\text B},\sum_{i}(\hat{K}^0_i+\hat{\sigma}^{\dagger}_i\hat{\sigma}_i)]=0$, which indicates a $U(1)$ symmetry of $\mathcal{M}_{\text JC}^{\text B}$. As shown in SM\cite{SM}, the $L$-dependence of $t^*$ for this model is similar to that of fermions and qubits. For a finite $L$, there is a crossover in the $L$-dependence of $t^*$ from $\log L$ to $L^2$ with an increasing $g/J$ for $JC$ mixing between bosons and qubits.  The eigenvalues of $\mathcal{M}_{\text JC}^{\text B}$ are identical to those of $\mathcal{M}_{\text JC}^{\text F}$ for $N=1$. The largest eigenvalues of $\mathcal{M}_{\text JC}^{\text B}$ for any finite $N$ become degenerate with those for $N=1$ with an increasing $L$ due to an emergent approximate symmetry of $\mathcal{M}_{\text JC}^{\text B}$.  The above features lead to the similarity between the fermionic and bosonic models with $JC$ mixing.

Next, we consider $R$ mixing between fermions or bosons and qubits. We start with the fermionic case having a finite-dimensional Hilbert space. The generating Hamiltonian in this case is (see Sec. IV of SM\cite{SM} for a derivation)
\bea
\mathcal{M}_{\text R}^{\text F}&=&\big(1-\frac{(2g^2+J^2)L}{2}\big)\mathds{1}_{\mathcal{N}_{\text R}^{\text F}}+\sum_{i=1}^L\big(\sum_{\nu}\frac{J^2}{2}\hat{\tau}_i^{\nu}\hat{\tau}_{i+1}^{\nu}\nn\\&+&g^2\hat{\tau}_i^{z}\hat{\sigma}_{i}^{z}\big)+\mathcal{O}(J^4,g^4), \label{mapHamRF}
\eea
which commutes with $\sum_{i=1}^L\hat{\tau}_i^{z}$ and $\hat{\sigma}_{j}^{z}$ for $j \in \{1,2,3,\dots,L\}$, indicating a global $U(1)$ symmetry for fermions and local $U(1)$ symmetry for each qubit. Interestingly, $\hat{H}(t)$ does not have a global $U(1)$ symmetry for $R$ mixing. The generating Hamiltonian for $R$  mixing does not have $SU(2)$ symmetry due to magnetic anisotropy created by coupling to the qubits in contrast to that in Eq.~\ref{mapHamJCF} for $JC$ mixing between fermions and qubits. The eigenvalues $\lambda_i$ of $\mathcal{M}_{\text R}^{\text F}$ can be determined by fixing $\sum_{i=1}^L\hat{\tau}_i^{z}$ and $\hat{\sigma}_{j}^{z}$ for $j \in \{1,2,3,\dots,L\}$ as these are good quantum numbers. The eigenvalues are doubly degenerate since $\mathcal{M}_{\text R}^{\text F}$ is invariant under $\prod_{i=1}^L\hat{\tau}_i^{x}\hat{\sigma}_{i}^{x}$, which implies a state obtained by flipping all the $\tau$ and $\sigma$ spins of an eigenstate of $\mathcal{M}_{\text R}^{\text F}$ is also an eigenstate with the same eigenvalue. The largest eigenvalue one of $\mathcal{M}_{\text R}^{\text F}$ is a state $|\lambda_0\ra$ in which all $\tau$ and $\sigma$ spins are polarized in $z$ direction.  The second largest eigenvalues of $\mathcal{M}_{\text R}^{\text F}$ are $L+1$ and $L$ fold degenerate, respectively, for $(g/J)^2<2/3$ and $(g/J)^2>2/3$ (see SM\cite{SM} for details). For $(g/J)^2<2/3$, the second largest eigenvalues are $1-2g^2$, which consist of $L$ eigenstates with anyone $\sigma$ spin being flipped in $|\lambda_0\ra$ and another superposition state with a single $\tau$ spin flipping in $|\lambda_0\ra$.  For $(g/J)^2>2/3$, the second largest eigenvalues are $1-4g^2-2J^2(1-\sqrt{1+4(g/J)^4})$, which are $L$ eigenstates with one $\tau$ spin flipping and one $\sigma$ spin being flipped in $|\lambda_0\ra$. Thus, the second largest eigenvalues for any $g/J$ are $L$ independent. So we get $t^* \propto \log L$ or $\log(L+1)$ for $R$ mixing between fermions and qubits. Such $L$-dependence of $t^*$ is similar to that in a periodically kicked transverse-field Ising model in \textcite{KosPRX2018} with local kicking terms. Interestingly, a similar $L$-scaling of $t^*$ can also be obtained for the $U(1)$-symmetry broken model explored in Ref.~\cite{RoyPRE2020} when the pairing $\Delta$ and tunneling $J$ strengths are the same. We can also get  $t^* \propto L^0$ of \textcite{RoyPRE2020} for arbitrary $\Delta$ and $J$ when $g$ is different/random for different qubits to lift the degeneracy in the second largest eigenvalues.     

Finally, we consider the mixture of bosons and qubits with $R$ mixing between them. The generating Hamiltonian for this case in the Trotter regime reads as (see Sec. V of SM\cite{SM} for a derivation)
\bea
\mathcal{M}_{\text R}^{\text B}&=&\big(1+\frac{J^2L}{2}\big)\mathds{1}_{\mathcal{N}_{\text R}^{\text B}}+\sum_{i=1}^L\Big(2J^2\big(\hat{K}^1_i\hat{K}^1_{i+1}+\hat{K}^2_{i}\hat{K}^2_{i+1}\nn\\&-&\hat{K}^0_i\hat{K}^0_{i+1}\big)+2g^2(\hat{K}^1_i\hat{\sigma}^x_i+\hat{K}^0_i)\Big)+\mathcal{O}(J^4,g^4), \label{mapHamRB}
\eea
which commutes with $\hat{\sigma}_{j}^{x}$ for $j \in \{1,2,3,\dots,L\}$. We could not calculate the spectrum of $\mathcal{M}_{\text R}^{\text B}$ analytically. Instead, we determine it numerically by varying $N_{max}$ for a fixed $L$ to get an estimate of $\lambda_1$ in the large $N_{max}$ limit. We use linear extrapolations in $1/N_{max}$ towards $1/N_{max}=0$ to find evidence for a gap between the largest and second largest eigenvalues, as shown in Sec. V of SM\cite{SM}. The second largest eigenvalues are also $L$-fold degenerate, suggesting a scaling of $t^* \propto \log L$. We remind here that a periodically kicked boson model without particle number conservation shows $t^*\propto \mathcal{O}(L^{\gamma}),\gamma=0.7 \pm 0.1$ \cite{RoyPRE2022}, which is sharply different from the present case of bosons and qubits without total number conservation.

We have analytically calculated the SFF in many-body quantum mixtures of fermions, bosons and qubits with periodically kicked spreading and mixing of species. Different types of mixing between species can drastically alter the timescale for the emergence of RMT behavior of $K(t)$ in quantum mixtures. We show how competition between mixing and hopping/spreading of species in $U(1)$-symmetric finite-size systems can lead to a logarithmic $L$ scaling of $t^*$, which has been predicted before only for $U(1)$-symmetry broken single-species models \cite{KosPRX2018, ChanPRL2018}. This finding is practical and vital as quantum mixtures of different species are abundant in nature as well as controlled experimental set-ups of cold atoms and photonic systems, and many of these systems are finite-sized. We further show the $t^*$ scaling for $R$ mixing of fermions and qubits is similar to those obtained for a single species of spin-1/2s or fermions. Finally, our results indicate that the $R$ mixing of species with different statistics (e.g., bosons and qubits) can lead to completely new features for the main species (e.g., bosons) with individual hopping. 

\textit{Acknowledgements:} We thank Prof. Toma$\check{\rm z}$ Prosen for many useful discussions.

\bibliography{bibliographyRMT}

\begin{thebibliography}{48}%
\makeatletter
\providecommand \@ifxundefined [1]{%
 \@ifx{#1\undefined}
}%
\providecommand \@ifnum [1]{%
 \ifnum #1\expandafter \@firstoftwo
 \else \expandafter \@secondoftwo
 \fi
}%
\providecommand \@ifx [1]{%
 \ifx #1\expandafter \@firstoftwo
 \else \expandafter \@secondoftwo
 \fi
}%
\providecommand \natexlab [1]{#1}%
\providecommand \enquote  [1]{``#1''}%
\providecommand \bibnamefont  [1]{#1}%
\providecommand \bibfnamefont [1]{#1}%
\providecommand \citenamefont [1]{#1}%
\providecommand \href@noop [0]{\@secondoftwo}%
\providecommand \href [0]{\begingroup \@sanitize@url \@href}%
\providecommand \@href[1]{\@@startlink{#1}\@@href}%
\providecommand \@@href[1]{\endgroup#1\@@endlink}%
\providecommand \@sanitize@url [0]{\catcode `\\12\catcode `\$12\catcode
  `\&12\catcode `\#12\catcode `\^12\catcode `\_12\catcode `\%12\relax}%
\providecommand \@@startlink[1]{}%
\providecommand \@@endlink[0]{}%
\providecommand \url  [0]{\begingroup\@sanitize@url \@url }%
\providecommand \@url [1]{\endgroup\@href {#1}{\urlprefix }}%
\providecommand \urlprefix  [0]{URL }%
\providecommand \Eprint [0]{\href }%
\providecommand \doibase [0]{https://doi.org/}%
\providecommand \selectlanguage [0]{\@gobble}%
\providecommand \bibinfo  [0]{\@secondoftwo}%
\providecommand \bibfield  [0]{\@secondoftwo}%
\providecommand \translation [1]{[#1]}%
\providecommand \BibitemOpen [0]{}%
\providecommand \bibitemStop [0]{}%
\providecommand \bibitemNoStop [0]{.\EOS\space}%
\providecommand \EOS [0]{\spacefactor3000\relax}%
\providecommand \BibitemShut  [1]{\csname bibitem#1\endcsname}%
\let\auto@bib@innerbib\@empty
\bibitem [{\citenamefont {Kos}\ \emph {et~al.}(2018)\citenamefont {Kos},
  \citenamefont {Ljubotina},\ and\ \citenamefont {Prosen}}]{KosPRX2018}%
  \BibitemOpen
  \bibfield  {author} {\bibinfo {author} {\bibfnamefont {P.}~\bibnamefont
  {Kos}}, \bibinfo {author} {\bibfnamefont {M.}~\bibnamefont {Ljubotina}},\
  and\ \bibinfo {author} {\bibfnamefont {T.}~\bibnamefont {Prosen}},\
  }\bibfield  {title} {\bibinfo {title} {Many-body quantum chaos: Analytic
  connection to random matrix theory},\ }\href
  {https://doi.org/10.1103/PhysRevX.8.021062} {\bibfield  {journal} {\bibinfo
  {journal} {Phys. Rev. X}\ }\textbf {\bibinfo {volume} {8}},\ \bibinfo {pages}
  {021062} (\bibinfo {year} {2018})}\BibitemShut {NoStop}%
\bibitem [{\citenamefont {Bertini}\ \emph {et~al.}(2018)\citenamefont
  {Bertini}, \citenamefont {Kos},\ and\ \citenamefont
  {Prosen}}]{BertiniPRL2018}%
  \BibitemOpen
  \bibfield  {author} {\bibinfo {author} {\bibfnamefont {B.}~\bibnamefont
  {Bertini}}, \bibinfo {author} {\bibfnamefont {P.}~\bibnamefont {Kos}},\ and\
  \bibinfo {author} {\bibfnamefont {T.}~\bibnamefont {Prosen}},\ }\bibfield
  {title} {\bibinfo {title} {Exact spectral form factor in a minimal model of
  many-body quantum chaos},\ }\href
  {https://doi.org/10.1103/PhysRevLett.121.264101} {\bibfield  {journal}
  {\bibinfo  {journal} {Phys. Rev. Lett.}\ }\textbf {\bibinfo {volume} {121}},\
  \bibinfo {pages} {264101} (\bibinfo {year} {2018})}\BibitemShut {NoStop}%
\bibitem [{\citenamefont {Chan}\ \emph
  {et~al.}(2018{\natexlab{a}})\citenamefont {Chan}, \citenamefont {De~Luca},\
  and\ \citenamefont {Chalker}}]{ChanPRX2018}%
  \BibitemOpen
  \bibfield  {author} {\bibinfo {author} {\bibfnamefont {A.}~\bibnamefont
  {Chan}}, \bibinfo {author} {\bibfnamefont {A.}~\bibnamefont {De~Luca}},\ and\
  \bibinfo {author} {\bibfnamefont {J.~T.}\ \bibnamefont {Chalker}},\
  }\bibfield  {title} {\bibinfo {title} {Solution of a minimal model for
  many-body quantum chaos},\ }\href {https://doi.org/10.1103/PhysRevX.8.041019}
  {\bibfield  {journal} {\bibinfo  {journal} {Phys. Rev. X}\ }\textbf {\bibinfo
  {volume} {8}},\ \bibinfo {pages} {041019} (\bibinfo {year}
  {2018}{\natexlab{a}})}\BibitemShut {NoStop}%
\bibitem [{\citenamefont {Chan}\ \emph
  {et~al.}(2018{\natexlab{b}})\citenamefont {Chan}, \citenamefont {De~Luca},\
  and\ \citenamefont {Chalker}}]{ChanPRL2018}%
  \BibitemOpen
  \bibfield  {author} {\bibinfo {author} {\bibfnamefont {A.}~\bibnamefont
  {Chan}}, \bibinfo {author} {\bibfnamefont {A.}~\bibnamefont {De~Luca}},\ and\
  \bibinfo {author} {\bibfnamefont {J.~T.}\ \bibnamefont {Chalker}},\
  }\bibfield  {title} {\bibinfo {title} {Spectral statistics in spatially
  extended chaotic quantum many-body systems},\ }\href
  {https://doi.org/10.1103/PhysRevLett.121.060601} {\bibfield  {journal}
  {\bibinfo  {journal} {Phys. Rev. Lett.}\ }\textbf {\bibinfo {volume} {121}},\
  \bibinfo {pages} {060601} (\bibinfo {year} {2018}{\natexlab{b}})}\BibitemShut
  {NoStop}%
\bibitem [{\citenamefont {Friedman}\ \emph {et~al.}(2019)\citenamefont
  {Friedman}, \citenamefont {Chan}, \citenamefont {De~Luca},\ and\
  \citenamefont {Chalker}}]{FriedmanPRL2019}%
  \BibitemOpen
  \bibfield  {author} {\bibinfo {author} {\bibfnamefont {A.~J.}\ \bibnamefont
  {Friedman}}, \bibinfo {author} {\bibfnamefont {A.}~\bibnamefont {Chan}},
  \bibinfo {author} {\bibfnamefont {A.}~\bibnamefont {De~Luca}},\ and\ \bibinfo
  {author} {\bibfnamefont {J.~T.}\ \bibnamefont {Chalker}},\ }\bibfield
  {title} {\bibinfo {title} {Spectral statistics and many-body quantum chaos
  with conserved charge},\ }\href
  {https://doi.org/10.1103/PhysRevLett.123.210603} {\bibfield  {journal}
  {\bibinfo  {journal} {Phys. Rev. Lett.}\ }\textbf {\bibinfo {volume} {123}},\
  \bibinfo {pages} {210603} (\bibinfo {year} {2019})}\BibitemShut {NoStop}%
\bibitem [{\citenamefont {Bertini}\ \emph {et~al.}(2019)\citenamefont
  {Bertini}, \citenamefont {Kos},\ and\ \citenamefont
  {Prosen}}]{Bertini2019PRL}%
  \BibitemOpen
  \bibfield  {author} {\bibinfo {author} {\bibfnamefont {B.}~\bibnamefont
  {Bertini}}, \bibinfo {author} {\bibfnamefont {P.}~\bibnamefont {Kos}},\ and\
  \bibinfo {author} {\bibfnamefont {T.}~\bibnamefont {Prosen}},\ }\bibfield
  {title} {\bibinfo {title} {Exact correlation functions for dual-unitary
  lattice models in $1+1$ dimensions},\ }\href
  {https://doi.org/10.1103/PhysRevLett.123.210601} {\bibfield  {journal}
  {\bibinfo  {journal} {Phys. Rev. Lett.}\ }\textbf {\bibinfo {volume} {123}},\
  \bibinfo {pages} {210601} (\bibinfo {year} {2019})}\BibitemShut {NoStop}%
\bibitem [{\citenamefont {Roy}\ and\ \citenamefont
  {Prosen}(2020)}]{RoyPRE2020}%
  \BibitemOpen
  \bibfield  {author} {\bibinfo {author} {\bibfnamefont {D.}~\bibnamefont
  {Roy}}\ and\ \bibinfo {author} {\bibfnamefont {T.}~\bibnamefont {Prosen}},\
  }\bibfield  {title} {\bibinfo {title} {Random matrix spectral form factor in
  kicked interacting fermionic chains},\ }\href
  {https://doi.org/10.1103/PhysRevE.102.060202} {\bibfield  {journal} {\bibinfo
   {journal} {Phys. Rev. E}\ }\textbf {\bibinfo {volume} {102}},\ \bibinfo
  {pages} {060202(R)} (\bibinfo {year} {2020})}\BibitemShut {NoStop}%
\bibitem [{\citenamefont {Bertini}\ \emph {et~al.}(2021)\citenamefont
  {Bertini}, \citenamefont {Kos},\ and\ \citenamefont
  {Prosen}}]{Bertini2021CMP}%
  \BibitemOpen
  \bibfield  {author} {\bibinfo {author} {\bibfnamefont {B.}~\bibnamefont
  {Bertini}}, \bibinfo {author} {\bibfnamefont {P.}~\bibnamefont {Kos}},\ and\
  \bibinfo {author} {\bibfnamefont {T.}~\bibnamefont {Prosen}},\ }\bibfield
  {title} {\bibinfo {title} {Random matrix spectral form factor of dual-unitary
  quantum circuits},\ }\href
  {https://doi.org/https://doi.org/10.1007/s00220-021-04139-2} {\bibfield
  {journal} {\bibinfo  {journal} {Commun. Math. Phys.}\ }\textbf {\bibinfo
  {volume} {387}},\ \bibinfo {pages} {597} (\bibinfo {year}
  {2021})}\BibitemShut {NoStop}%
\bibitem [{\citenamefont {Garratt}\ and\ \citenamefont
  {Chalker}(2021)}]{Garratt2021PRX}%
  \BibitemOpen
  \bibfield  {author} {\bibinfo {author} {\bibfnamefont {S.~J.}\ \bibnamefont
  {Garratt}}\ and\ \bibinfo {author} {\bibfnamefont {J.~T.}\ \bibnamefont
  {Chalker}},\ }\bibfield  {title} {\bibinfo {title} {Local pairing of feynman
  histories in many-body floquet models},\ }\href
  {https://doi.org/10.1103/PhysRevX.11.021051} {\bibfield  {journal} {\bibinfo
  {journal} {Phys. Rev. X}\ }\textbf {\bibinfo {volume} {11}},\ \bibinfo
  {pages} {021051} (\bibinfo {year} {2021})}\BibitemShut {NoStop}%
\bibitem [{\citenamefont {Li}\ \emph {et~al.}(2021)\citenamefont {Li},
  \citenamefont {Prosen},\ and\ \citenamefont {Chan}}]{LiPRL2021}%
  \BibitemOpen
  \bibfield  {author} {\bibinfo {author} {\bibfnamefont {J.}~\bibnamefont
  {Li}}, \bibinfo {author} {\bibfnamefont {T.}~\bibnamefont {Prosen}},\ and\
  \bibinfo {author} {\bibfnamefont {A.}~\bibnamefont {Chan}},\ }\bibfield
  {title} {\bibinfo {title} {Spectral statistics of non-{Hermitian} matrices
  and dissipative quantum chaos},\ }\href
  {https://doi.org/10.1103/PhysRevLett.127.170602} {\bibfield  {journal}
  {\bibinfo  {journal} {Phys. Rev. Lett.}\ }\textbf {\bibinfo {volume} {127}},\
  \bibinfo {pages} {170602} (\bibinfo {year} {2021})}\BibitemShut {NoStop}%
\bibitem [{\citenamefont {Kos}\ \emph {et~al.}(2021)\citenamefont {Kos},
  \citenamefont {Bertini},\ and\ \citenamefont {Prosen}}]{KosPRL2021}%
  \BibitemOpen
  \bibfield  {author} {\bibinfo {author} {\bibfnamefont {P.}~\bibnamefont
  {Kos}}, \bibinfo {author} {\bibfnamefont {B.}~\bibnamefont {Bertini}},\ and\
  \bibinfo {author} {\bibfnamefont {T.}~\bibnamefont {Prosen}},\ }\bibfield
  {title} {\bibinfo {title} {Chaos and ergodicity in extended quantum systems
  with noisy driving},\ }\href {https://doi.org/10.1103/PhysRevLett.126.190601}
  {\bibfield  {journal} {\bibinfo  {journal} {Phys. Rev. Lett.}\ }\textbf
  {\bibinfo {volume} {126}},\ \bibinfo {pages} {190601} (\bibinfo {year}
  {2021})}\BibitemShut {NoStop}%
\bibitem [{\citenamefont {Moudgalya}\ \emph {et~al.}(2021)\citenamefont
  {Moudgalya}, \citenamefont {Prem}, \citenamefont {Huse},\ and\ \citenamefont
  {Chan}}]{Moudgalya21}%
  \BibitemOpen
  \bibfield  {author} {\bibinfo {author} {\bibfnamefont {S.}~\bibnamefont
  {Moudgalya}}, \bibinfo {author} {\bibfnamefont {A.}~\bibnamefont {Prem}},
  \bibinfo {author} {\bibfnamefont {D.~A.}\ \bibnamefont {Huse}},\ and\
  \bibinfo {author} {\bibfnamefont {A.}~\bibnamefont {Chan}},\ }\bibfield
  {title} {\bibinfo {title} {Spectral statistics in constrained many-body
  quantum chaotic systems},\ }\href
  {https://doi.org/10.1103/PhysRevResearch.3.023176} {\bibfield  {journal}
  {\bibinfo  {journal} {Phys. Rev. Research}\ }\textbf {\bibinfo {volume}
  {3}},\ \bibinfo {pages} {023176} (\bibinfo {year} {2021})}\BibitemShut
  {NoStop}%
\bibitem [{\citenamefont {Roy}\ \emph {et~al.}(2022)\citenamefont {Roy},
  \citenamefont {Mishra},\ and\ \citenamefont {Prosen}}]{RoyPRE2022}%
  \BibitemOpen
  \bibfield  {author} {\bibinfo {author} {\bibfnamefont {D.}~\bibnamefont
  {Roy}}, \bibinfo {author} {\bibfnamefont {D.}~\bibnamefont {Mishra}},\ and\
  \bibinfo {author} {\bibfnamefont {T.}~\bibnamefont {Prosen}},\ }\bibfield
  {title} {\bibinfo {title} {Spectral form factor in a minimal bosonic model of
  many-body quantum chaos},\ }\href
  {https://doi.org/10.1103/PhysRevE.106.024208} {\bibfield  {journal} {\bibinfo
   {journal} {Phys. Rev. E}\ }\textbf {\bibinfo {volume} {106}},\ \bibinfo
  {pages} {024208} (\bibinfo {year} {2022})}\BibitemShut {NoStop}%
\bibitem [{\citenamefont {Winer}\ and\ \citenamefont
  {Swingle}(2022{\natexlab{a}})}]{Swingle22}%
  \BibitemOpen
  \bibfield  {author} {\bibinfo {author} {\bibfnamefont {M.}~\bibnamefont
  {Winer}}\ and\ \bibinfo {author} {\bibfnamefont {B.}~\bibnamefont
  {Swingle}},\ }\bibfield  {title} {\bibinfo {title} {The {Loschmidt} spectral
  form factor},\ }\href {https://doi.org/10.1007/JHEP10(2022)137} {\bibfield
  {journal} {\bibinfo  {journal} {J. High Energ. Phys.}\ }\textbf {\bibinfo
  {volume} {2022}},\ \bibinfo {pages} {137 (2022)}}\BibitemShut {NoStop}%
\bibitem [{\citenamefont {Winer}\ and\ \citenamefont
  {Swingle}(2022{\natexlab{b}})}]{Winer22}%
  \BibitemOpen
  \bibfield  {author} {\bibinfo {author} {\bibfnamefont {M.}~\bibnamefont
  {Winer}}\ and\ \bibinfo {author} {\bibfnamefont {B.}~\bibnamefont
  {Swingle}},\ }\bibfield  {title} {\bibinfo {title} {Hydrodynamic theory of
  the connected spectral form factor},\ }\href
  {https://doi.org/10.1103/PhysRevX.12.021009} {\bibfield  {journal} {\bibinfo
  {journal} {Phys. Rev. X}\ }\textbf {\bibinfo {volume} {12}},\ \bibinfo
  {pages} {021009} (\bibinfo {year} {2022}{\natexlab{b}})}\BibitemShut
  {NoStop}%
\bibitem [{\citenamefont {Liao}\ and\ \citenamefont
  {Galitski}(2022{\natexlab{a}})}]{Liao22}%
  \BibitemOpen
  \bibfield  {author} {\bibinfo {author} {\bibfnamefont {Y.}~\bibnamefont
  {Liao}}\ and\ \bibinfo {author} {\bibfnamefont {V.}~\bibnamefont
  {Galitski}},\ }\bibfield  {title} {\bibinfo {title} {Emergence of many-body
  quantum chaos via spontaneous breaking of unitarity},\ }\href
  {https://doi.org/10.1103/PhysRevB.105.L140202} {\bibfield  {journal}
  {\bibinfo  {journal} {Phys. Rev. B}\ }\textbf {\bibinfo {volume} {105}},\
  \bibinfo {pages} {L140202} (\bibinfo {year}
  {2022}{\natexlab{a}})}\BibitemShut {NoStop}%
\bibitem [{\citenamefont {Liao}\ and\ \citenamefont
  {Galitski}(2022{\natexlab{b}})}]{Liao2022}%
  \BibitemOpen
  \bibfield  {author} {\bibinfo {author} {\bibfnamefont {Y.}~\bibnamefont
  {Liao}}\ and\ \bibinfo {author} {\bibfnamefont {V.}~\bibnamefont
  {Galitski}},\ }\bibfield  {title} {\bibinfo {title} {Emergence of many-body
  quantum chaos via spontaneous breaking of unitarity},\ }\href
  {https://doi.org/10.1103/PhysRevB.105.L140202} {\bibfield  {journal}
  {\bibinfo  {journal} {Phys. Rev. B}\ }\textbf {\bibinfo {volume} {105}},\
  \bibinfo {pages} {L140202} (\bibinfo {year}
  {2022}{\natexlab{b}})}\BibitemShut {NoStop}%
\bibitem [{\citenamefont {Dag}\ \emph {et~al.}(2023)\citenamefont {Dag},
  \citenamefont {Mistakidis}, \citenamefont {Chan},\ and\ \citenamefont
  {Sadeghpour}}]{Dag23}%
  \BibitemOpen
  \bibfield  {author} {\bibinfo {author} {\bibfnamefont {C.}~\bibnamefont
  {Dag}}, \bibinfo {author} {\bibfnamefont {S.}~\bibnamefont {Mistakidis}},
  \bibinfo {author} {\bibfnamefont {A.}~\bibnamefont {Chan}},\ and\ \bibinfo
  {author} {\bibfnamefont {H.~R.}\ \bibnamefont {Sadeghpour}},\ }\bibfield
  {title} {\bibinfo {title} {Many-body quantum chaos in stroboscopically-driven
  cold atoms},\ }\href {https://doi.org/10.1038/s42005-023-01258-1} {\bibfield
  {journal} {\bibinfo  {journal} {Commun. Phys.}\ }\textbf {\bibinfo {volume}
  {6}},\ \bibinfo {pages} {136} (\bibinfo {year} {2023})}\BibitemShut {NoStop}%
\bibitem [{\citenamefont {Haake}(2001)}]{Haake2001}%
  \BibitemOpen
  \bibfield  {author} {\bibinfo {author} {\bibfnamefont {F.}~\bibnamefont
  {Haake}},\ }\href@noop {} {\emph {\bibinfo {title} {Quantum Signatures of
  Chaos, 2nd ed.}}}\ (\bibinfo  {publisher} {Springer, New York},\ \bibinfo
  {year} {2001})\BibitemShut {NoStop}%
\bibitem [{\citenamefont {Berry}(1985)}]{Berry1985}%
  \BibitemOpen
  \bibfield  {author} {\bibinfo {author} {\bibfnamefont {M.~V.}\ \bibnamefont
  {Berry}},\ }\bibfield  {title} {\bibinfo {title} {Semiclassical theory of
  spectral rigidity},\ }\href {https://doi.org/10.1098/rspa.1985.0078}
  {\bibfield  {journal} {\bibinfo  {journal} {Proc. R. Soc. Lond. A}\ }\textbf
  {\bibinfo {volume} {400}},\ \bibinfo {pages} {229} (\bibinfo {year}
  {1985})}\BibitemShut {NoStop}%
\bibitem [{\citenamefont {Bohigas}\ \emph {et~al.}(1984)\citenamefont
  {Bohigas}, \citenamefont {Giannoni},\ and\ \citenamefont
  {Schmit}}]{BohigasPRL1984}%
  \BibitemOpen
  \bibfield  {author} {\bibinfo {author} {\bibfnamefont {O.}~\bibnamefont
  {Bohigas}}, \bibinfo {author} {\bibfnamefont {M.~J.}\ \bibnamefont
  {Giannoni}},\ and\ \bibinfo {author} {\bibfnamefont {C.}~\bibnamefont
  {Schmit}},\ }\bibfield  {title} {\bibinfo {title} {Characterization of
  chaotic quantum spectra and universality of level fluctuation laws},\ }\href
  {https://doi.org/10.1103/PhysRevLett.52.1} {\bibfield  {journal} {\bibinfo
  {journal} {Phys. Rev. Lett.}\ }\textbf {\bibinfo {volume} {52}},\ \bibinfo
  {pages} {1} (\bibinfo {year} {1984})}\BibitemShut {NoStop}%
\bibitem [{\citenamefont {McDonald}\ and\ \citenamefont
  {Kaufman}(1979)}]{McDonaldPRL1979}%
  \BibitemOpen
  \bibfield  {author} {\bibinfo {author} {\bibfnamefont {S.~W.}\ \bibnamefont
  {McDonald}}\ and\ \bibinfo {author} {\bibfnamefont {A.~N.}\ \bibnamefont
  {Kaufman}},\ }\bibfield  {title} {\bibinfo {title} {Spectrum and
  eigenfunctions for a {Hamiltonian} with stochastic trajectories},\ }\href
  {https://doi.org/10.1103/PhysRevLett.42.1189} {\bibfield  {journal} {\bibinfo
   {journal} {Phys. Rev. Lett.}\ }\textbf {\bibinfo {volume} {42}},\ \bibinfo
  {pages} {1189} (\bibinfo {year} {1979})}\BibitemShut {NoStop}%
\bibitem [{\citenamefont {Casati}\ \emph {et~al.}(1980)\citenamefont {Casati},
  \citenamefont {Valz-Gris},\ and\ \citenamefont {Guarnieri}}]{Casati1980}%
  \BibitemOpen
  \bibfield  {author} {\bibinfo {author} {\bibfnamefont {G.}~\bibnamefont
  {Casati}}, \bibinfo {author} {\bibfnamefont {F.}~\bibnamefont {Valz-Gris}},\
  and\ \bibinfo {author} {\bibfnamefont {I.}~\bibnamefont {Guarnieri}},\
  }\bibfield  {title} {\bibinfo {title} {On the connection between quantization
  of nonintegrable systems and statistical theory of spectra},\ }\href
  {https://doi.org/10.1007/BF02798790} {\bibfield  {journal} {\bibinfo
  {journal} {Lett. Nuovo Cimento}\ }\textbf {\bibinfo {volume} {28}},\ \bibinfo
  {pages} {279} (\bibinfo {year} {1980})}\BibitemShut {NoStop}%
\bibitem [{\citenamefont {Berry}\ and\ \citenamefont
  {Tabor}(1977)}]{Berry1977}%
  \BibitemOpen
  \bibfield  {author} {\bibinfo {author} {\bibfnamefont {M.~V.}\ \bibnamefont
  {Berry}}\ and\ \bibinfo {author} {\bibfnamefont {M.}~\bibnamefont {Tabor}},\
  }\bibfield  {title} {\bibinfo {title} {Level clustering in the regular
  spectrum},\ }\href {https://doi.org/10.1098/rspa.1977.0140} {\bibfield
  {journal} {\bibinfo  {journal} {Proc. R. Soc. Lond. A}\ }\textbf {\bibinfo
  {volume} {356}},\ \bibinfo {pages} {375} (\bibinfo {year}
  {1977})}\BibitemShut {NoStop}%
\bibitem [{\citenamefont {Berry}(1981)}]{Berry1981}%
  \BibitemOpen
  \bibfield  {author} {\bibinfo {author} {\bibfnamefont {M.~V.}\ \bibnamefont
  {Berry}},\ }\bibfield  {title} {\bibinfo {title} {Quantizing a classically
  ergodic system: {Sinai's} billiard and the {KKR} method},\ }\href
  {https://doi.org/10.1016/0003-4916(81)90189-5} {\bibfield  {journal}
  {\bibinfo  {journal} {Ann. Phys.}\ }\textbf {\bibinfo {volume} {131}},\
  \bibinfo {pages} {163} (\bibinfo {year} {1981})}\BibitemShut {NoStop}%
\bibitem [{\citenamefont {Sieber}\ and\ \citenamefont
  {Richter}(2001)}]{Sieber2001}%
  \BibitemOpen
  \bibfield  {author} {\bibinfo {author} {\bibfnamefont {M.}~\bibnamefont
  {Sieber}}\ and\ \bibinfo {author} {\bibfnamefont {K.}~\bibnamefont
  {Richter}},\ }\bibfield  {title} {\bibinfo {title} {Correlations between
  periodic orbits and their role in spectral statistics},\ }\href
  {https://doi.org/10.1238/physica.topical.090a00128} {\bibfield  {journal}
  {\bibinfo  {journal} {Phys. Scr.}\ }\textbf {\bibinfo {volume} {T90}},\
  \bibinfo {pages} {128} (\bibinfo {year} {2001})}\BibitemShut {NoStop}%
\bibitem [{\citenamefont {Sieber}(2002)}]{Sieber2002}%
  \BibitemOpen
  \bibfield  {author} {\bibinfo {author} {\bibfnamefont {M.}~\bibnamefont
  {Sieber}},\ }\bibfield  {title} {\bibinfo {title} {Leading off-diagonal
  approximation for the spectral form factor for uniformly hyperbolic
  systems},\ }\href {https://doi.org/10.1088/0305-4470/35/42/104} {\bibfield
  {journal} {\bibinfo  {journal} {J. Phys. A}\ }\textbf {\bibinfo {volume}
  {35}},\ \bibinfo {pages} {L613} (\bibinfo {year} {2002})}\BibitemShut
  {NoStop}%
\bibitem [{\citenamefont {M\"uller}\ \emph {et~al.}(2004)\citenamefont
  {M\"uller}, \citenamefont {Heusler}, \citenamefont {Braun}, \citenamefont
  {Haake},\ and\ \citenamefont {Altland}}]{MullerPRL2004}%
  \BibitemOpen
  \bibfield  {author} {\bibinfo {author} {\bibfnamefont {S.}~\bibnamefont
  {M\"uller}}, \bibinfo {author} {\bibfnamefont {S.}~\bibnamefont {Heusler}},
  \bibinfo {author} {\bibfnamefont {P.}~\bibnamefont {Braun}}, \bibinfo
  {author} {\bibfnamefont {F.}~\bibnamefont {Haake}},\ and\ \bibinfo {author}
  {\bibfnamefont {A.}~\bibnamefont {Altland}},\ }\bibfield  {title} {\bibinfo
  {title} {Semiclassical foundation of universality in quantum chaos},\ }\href
  {https://doi.org/10.1103/PhysRevLett.93.014103} {\bibfield  {journal}
  {\bibinfo  {journal} {Phys. Rev. Lett.}\ }\textbf {\bibinfo {volume} {93}},\
  \bibinfo {pages} {014103} (\bibinfo {year} {2004})}\BibitemShut {NoStop}%
\bibitem [{\citenamefont {M\"uller}\ \emph {et~al.}(2005)\citenamefont
  {M\"uller}, \citenamefont {Heusler}, \citenamefont {Braun}, \citenamefont
  {Haake},\ and\ \citenamefont {Altland}}]{MullerPRE2005}%
  \BibitemOpen
  \bibfield  {author} {\bibinfo {author} {\bibfnamefont {S.}~\bibnamefont
  {M\"uller}}, \bibinfo {author} {\bibfnamefont {S.}~\bibnamefont {Heusler}},
  \bibinfo {author} {\bibfnamefont {P.}~\bibnamefont {Braun}}, \bibinfo
  {author} {\bibfnamefont {F.}~\bibnamefont {Haake}},\ and\ \bibinfo {author}
  {\bibfnamefont {A.}~\bibnamefont {Altland}},\ }\bibfield  {title} {\bibinfo
  {title} {Periodic-orbit theory of universality in quantum chaos},\ }\href
  {https://doi.org/10.1103/PhysRevE.72.046207} {\bibfield  {journal} {\bibinfo
  {journal} {Phys. Rev. E}\ }\textbf {\bibinfo {volume} {72}},\ \bibinfo
  {pages} {046207} (\bibinfo {year} {2005})}\BibitemShut {NoStop}%
\bibitem [{\citenamefont {Ch\'avez-Carlos}\ \emph {et~al.}(2019)\citenamefont
  {Ch\'avez-Carlos}, \citenamefont {L\'opez-del Carpio}, \citenamefont
  {Bastarrachea-Magnani}, \citenamefont {Str\'ansk\'y}, \citenamefont
  {Lerma-Hern\'andez}, \citenamefont {Santos},\ and\ \citenamefont
  {Hirsch}}]{CarlosPRL2019}%
  \BibitemOpen
  \bibfield  {author} {\bibinfo {author} {\bibfnamefont {J.}~\bibnamefont
  {Ch\'avez-Carlos}}, \bibinfo {author} {\bibfnamefont {B.}~\bibnamefont
  {L\'opez-del Carpio}}, \bibinfo {author} {\bibfnamefont {M.~A.}\ \bibnamefont
  {Bastarrachea-Magnani}}, \bibinfo {author} {\bibfnamefont {P.}~\bibnamefont
  {Str\'ansk\'y}}, \bibinfo {author} {\bibfnamefont {S.}~\bibnamefont
  {Lerma-Hern\'andez}}, \bibinfo {author} {\bibfnamefont {L.~F.}\ \bibnamefont
  {Santos}},\ and\ \bibinfo {author} {\bibfnamefont {J.~G.}\ \bibnamefont
  {Hirsch}},\ }\bibfield  {title} {\bibinfo {title} {Quantum and classical
  {Lyapunov} exponents in atom-field interaction systems},\ }\href
  {https://doi.org/10.1103/PhysRevLett.122.024101} {\bibfield  {journal}
  {\bibinfo  {journal} {Phys. Rev. Lett.}\ }\textbf {\bibinfo {volume} {122}},\
  \bibinfo {pages} {024101} (\bibinfo {year} {2019})}\BibitemShut {NoStop}%
\bibitem [{\citenamefont {Jaynes}\ and\ \citenamefont
  {Cummings}(1963)}]{Jaynes1963}%
  \BibitemOpen
  \bibfield  {author} {\bibinfo {author} {\bibfnamefont {E.}~\bibnamefont
  {Jaynes}}\ and\ \bibinfo {author} {\bibfnamefont {F.}~\bibnamefont
  {Cummings}},\ }\bibfield  {title} {\bibinfo {title} {Chaos and ergodicity in
  extended quantum systems with noisy driving},\ }\href
  {https://doi.org/10.1109/PROC.1963.1664} {\bibfield  {journal} {\bibinfo
  {journal} {Proc. IEEE.}\ }\textbf {\bibinfo {volume} {51}},\ \bibinfo {pages}
  {89} (\bibinfo {year} {1963})}\BibitemShut {NoStop}%
\bibitem [{\citenamefont {Scully}\ and\ \citenamefont
  {Zubairy}(1997)}]{scully1997quantum}%
  \BibitemOpen
  \bibfield  {author} {\bibinfo {author} {\bibfnamefont {M.}~\bibnamefont
  {Scully}}\ and\ \bibinfo {author} {\bibfnamefont {M.}~\bibnamefont
  {Zubairy}},\ }\href {https://books.google.co.in/books?id=9lkgAwAAQBAJ} {\emph
  {\bibinfo {title} {Quantum Optics}}}\ (\bibinfo  {publisher} {Cambridge
  University Press},\ \bibinfo {year} {1997})\BibitemShut {NoStop}%
\bibitem [{\citenamefont {Hartmann}\ \emph {et~al.}(2006)\citenamefont
  {Hartmann}, \citenamefont {Brand{\~{a}}o},\ and\ \citenamefont
  {Plenio}}]{Hartmann_2006}%
  \BibitemOpen
  \bibfield  {author} {\bibinfo {author} {\bibfnamefont {M.~J.}\ \bibnamefont
  {Hartmann}}, \bibinfo {author} {\bibfnamefont {F.~G. S.~L.}\ \bibnamefont
  {Brand{\~{a}}o}},\ and\ \bibinfo {author} {\bibfnamefont {M.~B.}\
  \bibnamefont {Plenio}},\ }\bibfield  {title} {\bibinfo {title} {Strongly
  interacting polaritons in coupled arrays of cavities},\ }\href
  {https://doi.org/10.1038/nphys462} {\bibfield  {journal} {\bibinfo  {journal}
  {Nature Physics}\ }\textbf {\bibinfo {volume} {2}},\ \bibinfo {pages} {849}
  (\bibinfo {year} {2006})}\BibitemShut {NoStop}%
\bibitem [{\citenamefont {Angelakis}\ \emph {et~al.}(2007)\citenamefont
  {Angelakis}, \citenamefont {Santos},\ and\ \citenamefont
  {Bose}}]{Angelakis2007}%
  \BibitemOpen
  \bibfield  {author} {\bibinfo {author} {\bibfnamefont {D.~G.}\ \bibnamefont
  {Angelakis}}, \bibinfo {author} {\bibfnamefont {M.~F.}\ \bibnamefont
  {Santos}},\ and\ \bibinfo {author} {\bibfnamefont {S.}~\bibnamefont {Bose}},\
  }\bibfield  {title} {\bibinfo {title} {Photon-blockade-induced {Mott}
  transitions and {$XY$} spin models in coupled cavity arrays},\ }\href
  {https://doi.org/10.1103/PhysRevA.76.031805} {\bibfield  {journal} {\bibinfo
  {journal} {Phys. Rev. A}\ }\textbf {\bibinfo {volume} {76}},\ \bibinfo
  {pages} {031805} (\bibinfo {year} {2007})}\BibitemShut {NoStop}%
\bibitem [{\citenamefont {Roy}\ \emph {et~al.}(2017)\citenamefont {Roy},
  \citenamefont {Wilson},\ and\ \citenamefont {Firstenberg}}]{RoyRMP2017}%
  \BibitemOpen
  \bibfield  {author} {\bibinfo {author} {\bibfnamefont {D.}~\bibnamefont
  {Roy}}, \bibinfo {author} {\bibfnamefont {C.~M.}\ \bibnamefont {Wilson}},\
  and\ \bibinfo {author} {\bibfnamefont {O.}~\bibnamefont {Firstenberg}},\
  }\bibfield  {title} {\bibinfo {title} {Colloquium: Strongly interacting
  photons in one-dimensional continuum},\ }\href
  {https://doi.org/10.1103/RevModPhys.89.021001} {\bibfield  {journal}
  {\bibinfo  {journal} {Rev. Mod. Phys.}\ }\textbf {\bibinfo {volume} {89}},\
  \bibinfo {pages} {021001} (\bibinfo {year} {2017})}\BibitemShut {NoStop}%
\bibitem [{\citenamefont {Rabi}(1937)}]{Rabi1937}%
  \BibitemOpen
  \bibfield  {author} {\bibinfo {author} {\bibfnamefont {I.~I.}\ \bibnamefont
  {Rabi}},\ }\bibfield  {title} {\bibinfo {title} {Space quantization in a
  gyrating magnetic field},\ }\href {https://doi.org/10.1103/PhysRev.51.652}
  {\bibfield  {journal} {\bibinfo  {journal} {Phys. Rev.}\ }\textbf {\bibinfo
  {volume} {51}},\ \bibinfo {pages} {652} (\bibinfo {year} {1937})}\BibitemShut
  {NoStop}%
\bibitem [{\citenamefont {Xie}\ \emph {et~al.}(2017)\citenamefont {Xie},
  \citenamefont {Zhong}, \citenamefont {Batchelor},\ and\ \citenamefont
  {Lee}}]{Xie2017}%
  \BibitemOpen
  \bibfield  {author} {\bibinfo {author} {\bibfnamefont {Q.}~\bibnamefont
  {Xie}}, \bibinfo {author} {\bibfnamefont {H.}~\bibnamefont {Zhong}}, \bibinfo
  {author} {\bibfnamefont {M.~T.}\ \bibnamefont {Batchelor}},\ and\ \bibinfo
  {author} {\bibfnamefont {C.}~\bibnamefont {Lee}},\ }\bibfield  {title}
  {\bibinfo {title} {The quantum {Rabi} model: solution and dynamics},\ }\href
  {https://doi.org/10.1088/1751-8121/aa5a65} {\bibfield  {journal} {\bibinfo
  {journal} {J. Phys. A}\ }\textbf {\bibinfo {volume} {50}},\ \bibinfo {pages}
  {113001} (\bibinfo {year} {2017})}\BibitemShut {NoStop}%
\bibitem [{\citenamefont {Agarwal}\ \emph {et~al.}(2023)\citenamefont
  {Agarwal}, \citenamefont {Sahu},\ and\ \citenamefont {Xu}}]{Agarwal2023}%
  \BibitemOpen
  \bibfield  {author} {\bibinfo {author} {\bibfnamefont {L.}~\bibnamefont
  {Agarwal}}, \bibinfo {author} {\bibfnamefont {S.}~\bibnamefont {Sahu}},\ and\
  \bibinfo {author} {\bibfnamefont {S.}~\bibnamefont {Xu}},\ }\bibfield
  {title} {\bibinfo {title} {Charge transport, information scrambling and
  quantum operator-coherence in a many-body system with {U(1)} symmetry},\
  }\href {https://doi.org/10.1007/JHEP05(2023)037} {\bibfield  {journal}
  {\bibinfo  {journal} {J. High Energ. Phys.}\ }\textbf {\bibinfo {volume}
  {2023}},\ \bibinfo {pages} {37 (2023)}}\BibitemShut {NoStop}%
\bibitem [{\citenamefont {Gharibyan}\ \emph {et~al.}(2018)\citenamefont
  {Gharibyan}, \citenamefont {Hanada}, \citenamefont {Shenker},\ and\
  \citenamefont {Tezuka}}]{Gharibyan2018}%
  \BibitemOpen
  \bibfield  {author} {\bibinfo {author} {\bibfnamefont {H.}~\bibnamefont
  {Gharibyan}}, \bibinfo {author} {\bibfnamefont {M.}~\bibnamefont {Hanada}},
  \bibinfo {author} {\bibfnamefont {S.~H.}\ \bibnamefont {Shenker}},\ and\
  \bibinfo {author} {\bibfnamefont {M.}~\bibnamefont {Tezuka}},\ }\bibfield
  {title} {\bibinfo {title} {Onset of random matrix behavior in scrambling
  systems},\ }\href {https://doi.org/10.1007/JHEP07(2018)124} {\bibfield
  {journal} {\bibinfo  {journal} {J. High Energ. Phys.}\ }\textbf {\bibinfo
  {volume} {2018}},\ \bibinfo {pages} {124 (2018)}}\BibitemShut {NoStop}%
\bibitem [{\citenamefont {Islam}\ \emph {et~al.}(2015)\citenamefont {Islam},
  \citenamefont {Ma}, \citenamefont {Preiss}, \citenamefont {Eric~Tai},
  \citenamefont {Lukin}, \citenamefont {Rispoli},\ and\ \citenamefont
  {Greiner}}]{islam2015measuring}%
  \BibitemOpen
  \bibfield  {author} {\bibinfo {author} {\bibfnamefont {R.}~\bibnamefont
  {Islam}}, \bibinfo {author} {\bibfnamefont {R.}~\bibnamefont {Ma}}, \bibinfo
  {author} {\bibfnamefont {P.~M.}\ \bibnamefont {Preiss}}, \bibinfo {author}
  {\bibfnamefont {M.}~\bibnamefont {Eric~Tai}}, \bibinfo {author}
  {\bibfnamefont {A.}~\bibnamefont {Lukin}}, \bibinfo {author} {\bibfnamefont
  {M.}~\bibnamefont {Rispoli}},\ and\ \bibinfo {author} {\bibfnamefont
  {M.}~\bibnamefont {Greiner}},\ }\bibfield  {title} {\bibinfo {title}
  {Measuring entanglement entropy in a quantum many-body system},\ }\href
  {https://doi.org/10.1038/nature15750} {\bibfield  {journal} {\bibinfo
  {journal} {Nature}\ }\textbf {\bibinfo {volume} {528}},\ \bibinfo {pages}
  {77} (\bibinfo {year} {2015})}\BibitemShut {NoStop}%
\bibitem [{\citenamefont {Kaufman}\ \emph {et~al.}(2016)\citenamefont
  {Kaufman}, \citenamefont {Tai}, \citenamefont {Lukin}, \citenamefont
  {Rispoli}, \citenamefont {Schittko}, \citenamefont {Preiss},\ and\
  \citenamefont {Greiner}}]{kaufman2016quantum}%
  \BibitemOpen
  \bibfield  {author} {\bibinfo {author} {\bibfnamefont {A.~M.}\ \bibnamefont
  {Kaufman}}, \bibinfo {author} {\bibfnamefont {M.~E.}\ \bibnamefont {Tai}},
  \bibinfo {author} {\bibfnamefont {A.}~\bibnamefont {Lukin}}, \bibinfo
  {author} {\bibfnamefont {M.}~\bibnamefont {Rispoli}}, \bibinfo {author}
  {\bibfnamefont {R.}~\bibnamefont {Schittko}}, \bibinfo {author}
  {\bibfnamefont {P.~M.}\ \bibnamefont {Preiss}},\ and\ \bibinfo {author}
  {\bibfnamefont {M.}~\bibnamefont {Greiner}},\ }\bibfield  {title} {\bibinfo
  {title} {Quantum thermalization through entanglement in an isolated many-body
  system},\ }\href {https://doi.org/10.1126/science.aaf6725} {\bibfield
  {journal} {\bibinfo  {journal} {Science}\ }\textbf {\bibinfo {volume}
  {353}},\ \bibinfo {pages} {794} (\bibinfo {year} {2016})}\BibitemShut
  {NoStop}%
\bibitem [{\citenamefont {Bernien}\ \emph {et~al.}(2017)\citenamefont
  {Bernien}, \citenamefont {Schwartz}, \citenamefont {Keesling}, \citenamefont
  {Levine}, \citenamefont {Omran}, \citenamefont {Pichler}, \citenamefont
  {Choi}, \citenamefont {Zibrov}, \citenamefont {Endres}, \citenamefont
  {Greiner} \emph {et~al.}}]{bernien2017probing}%
  \BibitemOpen
  \bibfield  {author} {\bibinfo {author} {\bibfnamefont {H.}~\bibnamefont
  {Bernien}}, \bibinfo {author} {\bibfnamefont {S.}~\bibnamefont {Schwartz}},
  \bibinfo {author} {\bibfnamefont {A.}~\bibnamefont {Keesling}}, \bibinfo
  {author} {\bibfnamefont {H.}~\bibnamefont {Levine}}, \bibinfo {author}
  {\bibfnamefont {A.}~\bibnamefont {Omran}}, \bibinfo {author} {\bibfnamefont
  {H.}~\bibnamefont {Pichler}}, \bibinfo {author} {\bibfnamefont
  {S.}~\bibnamefont {Choi}}, \bibinfo {author} {\bibfnamefont {A.~S.}\
  \bibnamefont {Zibrov}}, \bibinfo {author} {\bibfnamefont {M.}~\bibnamefont
  {Endres}}, \bibinfo {author} {\bibfnamefont {M.}~\bibnamefont {Greiner}},
  \emph {et~al.},\ }\bibfield  {title} {\bibinfo {title} {Probing many-body
  dynamics on a 51-atom quantum simulator},\ }\href
  {https://doi.org/10.1038/nature24622} {\bibfield  {journal} {\bibinfo
  {journal} {Nature}\ }\textbf {\bibinfo {volume} {551}},\ \bibinfo {pages}
  {579} (\bibinfo {year} {2017})}\BibitemShut {NoStop}%
\bibitem [{\citenamefont {Cronenberger}\ \emph {et~al.}(2019)\citenamefont
  {Cronenberger}, \citenamefont {Abbas}, \citenamefont {Scalbert},\ and\
  \citenamefont {Boukari}}]{Cronenberger2019}%
  \BibitemOpen
  \bibfield  {author} {\bibinfo {author} {\bibfnamefont {S.}~\bibnamefont
  {Cronenberger}}, \bibinfo {author} {\bibfnamefont {C.}~\bibnamefont {Abbas}},
  \bibinfo {author} {\bibfnamefont {D.}~\bibnamefont {Scalbert}},\ and\
  \bibinfo {author} {\bibfnamefont {H.}~\bibnamefont {Boukari}},\ }\bibfield
  {title} {\bibinfo {title} {Spatiotemporal spin noise spectroscopy},\ }\href
  {https://doi.org/10.1103/PhysRevLett.123.017401} {\bibfield  {journal}
  {\bibinfo  {journal} {Phys. Rev. Lett.}\ }\textbf {\bibinfo {volume} {123}},\
  \bibinfo {pages} {017401} (\bibinfo {year} {2019})}\BibitemShut {NoStop}%
\bibitem [{\citenamefont {Swar}\ \emph {et~al.}(2021)\citenamefont {Swar},
  \citenamefont {Roy}, \citenamefont {Bhar}, \citenamefont {Roy},\ and\
  \citenamefont {Chaudhuri}}]{Swar2021}%
  \BibitemOpen
  \bibfield  {author} {\bibinfo {author} {\bibfnamefont {M.}~\bibnamefont
  {Swar}}, \bibinfo {author} {\bibfnamefont {D.}~\bibnamefont {Roy}}, \bibinfo
  {author} {\bibfnamefont {S.}~\bibnamefont {Bhar}}, \bibinfo {author}
  {\bibfnamefont {S.}~\bibnamefont {Roy}},\ and\ \bibinfo {author}
  {\bibfnamefont {S.}~\bibnamefont {Chaudhuri}},\ }\bibfield  {title} {\bibinfo
  {title} {Detection of spin coherence in cold atoms via {Faraday} rotation
  fluctuations},\ }\href {https://doi.org/10.1103/PhysRevResearch.3.043171}
  {\bibfield  {journal} {\bibinfo  {journal} {Phys. Rev. Res.}\ }\textbf
  {\bibinfo {volume} {3}},\ \bibinfo {pages} {043171} (\bibinfo {year}
  {2021})}\BibitemShut {NoStop}%
\bibitem [{\citenamefont {Joshi}\ \emph {et~al.}(2022)\citenamefont {Joshi},
  \citenamefont {Elben}, \citenamefont {Vikram}, \citenamefont {Vermersch},
  \citenamefont {Galitski},\ and\ \citenamefont {Zoller}}]{Joshi2022}%
  \BibitemOpen
  \bibfield  {author} {\bibinfo {author} {\bibfnamefont {L.~K.}\ \bibnamefont
  {Joshi}}, \bibinfo {author} {\bibfnamefont {A.}~\bibnamefont {Elben}},
  \bibinfo {author} {\bibfnamefont {A.}~\bibnamefont {Vikram}}, \bibinfo
  {author} {\bibfnamefont {B.}~\bibnamefont {Vermersch}}, \bibinfo {author}
  {\bibfnamefont {V.}~\bibnamefont {Galitski}},\ and\ \bibinfo {author}
  {\bibfnamefont {P.}~\bibnamefont {Zoller}},\ }\bibfield  {title} {\bibinfo
  {title} {Probing many-body quantum chaos with quantum simulators},\ }\href
  {https://doi.org/10.1103/PhysRevX.12.011018} {\bibfield  {journal} {\bibinfo
  {journal} {Phys. Rev. X}\ }\textbf {\bibinfo {volume} {12}},\ \bibinfo
  {pages} {011018} (\bibinfo {year} {2022})}\BibitemShut {NoStop}%
\bibitem [{SM()}]{SM}%
  \BibitemOpen
  \href@noop {} {\bibinfo {title} {See {Supplemental} {Materials (SM)} for the
  derivation of spectral form factor and generating {Hamiltonians} for
  different mixing, and the details of spectral properties of these
  {Hamiltonians} in the {Trotter} regime.}}\BibitemShut {Stop}%
\bibitem [{\citenamefont {Roy}\ \emph {et~al.}(2015)\citenamefont {Roy},
  \citenamefont {Singh},\ and\ \citenamefont {Moessner}}]{Roy2015}%
  \BibitemOpen
  \bibfield  {author} {\bibinfo {author} {\bibfnamefont {D.}~\bibnamefont
  {Roy}}, \bibinfo {author} {\bibfnamefont {R.}~\bibnamefont {Singh}},\ and\
  \bibinfo {author} {\bibfnamefont {R.}~\bibnamefont {Moessner}},\ }\bibfield
  {title} {\bibinfo {title} {Probing many-body localization by spin noise
  spectroscopy},\ }\href {https://doi.org/10.1103/PhysRevB.92.180205}
  {\bibfield  {journal} {\bibinfo  {journal} {Phys. Rev. B}\ }\textbf {\bibinfo
  {volume} {92}},\ \bibinfo {pages} {180205(R)} (\bibinfo {year}
  {2015})}\BibitemShut {NoStop}%
\bibitem [{\citenamefont {Economou}(2006)}]{economou2006green}%
  \BibitemOpen
  \bibfield  {author} {\bibinfo {author} {\bibfnamefont {E.~N.}\ \bibnamefont
  {Economou}},\ }\href@noop {} {\emph {\bibinfo {title} {Green's functions in
  quantum physics}}},\ Vol.~\bibinfo {volume} {7}\ (\bibinfo  {publisher}
  {Springer Science \& Business Media},\ \bibinfo {year} {2006})\BibitemShut
  {NoStop}%
\end{thebibliography}%

\onecolumngrid
\vspace{15cm}
\begin{center}
{\large \bf{Supplementary Material for ``Many-Body Quantum Chaos in Mixtures of Multiple Species''}} \\
\vspace{2mm}
{Vijay Kumar and Dibyendu Roy} \\
\vspace{2mm}
{Raman Research Institute, Bangalore 560080, India}
\end{center}
\vspace{1cm}
\twocolumngrid
\setcounter{figure}{0}
\renewcommand\thefigure{S\arabic{figure}}
\setcounter{equation}{0}
\renewcommand\theequation{S\arabic{equation}}
\section{Spectral Form Factor for Periodically Kicked Systems}\label{SFF}
The spectral form factor (SFF) is defined as the Fourier transform of the two-point correlation function of spectral density. For a periodically kicked system, the spectral density is, $\rho(\varphi)=\left(2\pi/\mathcal{N}\right)\sum_{n=1}^\mathcal{N}\delta(\varphi-\varphi_n)$, where $\varphi_n$'s are the eigenphases of the Floquet propagator, $\hat{U}$, and $\mathcal{N}$ is the dimension of the system's Hilbert space. The prefactor, $2\pi/\mathcal{N}$, is chosen such that the averaged spectral density is normalized to unity, i.e.,
\begin{align}
\label{S1}
\langle \rho(\varphi)\rangle_\varphi=\int_0^{2\pi}d\varphi\: \rho(\varphi)=1.
\end{align}
We define a two-point correlation function of $\rho(\varphi)$ as
\begin{align}
\label{S2}
R(\vartheta)=\langle\rho(\varphi+\vartheta/2)\rho(\varphi-\vartheta/2)\rangle_\varphi-\langle\rho(\varphi)\rangle_\varphi^2.
\end{align}
We can then write the SFF as
\begin{align}
\label{S3}
K(t)=\frac{\mathcal{N}^2}{2\pi}\int_0^{2\pi}d\vartheta R(\vartheta)e^{-i\vartheta t}=(\text{tr}\hat{U}^t)(\text{tr}\hat{U}^{-t})-\mathcal{N}^2\delta_{t,0}.
\end{align}
Since $K(t)$ in Eq.~\ref{S3} is not self-averaging over disorder in onsite energy and transition frequency, we average it over different disorder realizations of them.
\begin{align}
\label{S4}
K(t)=\langle (\text{tr}\hat{U}^t)(\text{tr}\hat{U}^{-t})\rangle-\mathcal{N}^2\delta_{t,0},
\end{align}
where $\langle...\rangle$ represent averaging over disorder realizations. In the main paper, we consider periodically kicked quantum mixtures of multiple species whose Hamiltonian reads
\begin{align}
\label{S5}
\hat{H}(t)&=\hat{H}_0+\hat{H}_{\text{JC/R}}\sum_{m\in \mathcal{Z}}\delta(t-m),
\end{align}
where $\hat{H}_0$ is the base Hamiltonian and $\hat{H}_{\text{JC/R}}$ is the driving Hamiltonian. These Hamiltonians are given in the main paper. The Floquet propagator over the unit cycle of kicking is
\begin{align}
\label{S6}
\hat{U}&=\mathcal{T}e^{-i\int_0^1dt \hat{H}(t)}=\lim_{\epsilon\rightarrow 0}e^{-i\int_\epsilon^{1+\epsilon}dt \hat{H}(t)}=\hat{V}\hat{W},
\end{align}
where $\hat{W}=e^{-i\hat{H}_0},\hat{V}=e^{-i\hat{H}_{\text{JC/R}}}$. To proceed with the calculation of the SFF in Eq.~\ref{S4}, we choose the occupation number basis, $|\underline{n\sigma}\rangle\equiv |n_1,...,n_L\rangle\otimes|\sigma_1,...,\sigma_L\rangle$, which are the eigenbasis of $H_0$ \cite{RoyPRE2020,RoyPRE2022}. Here, $n_j,\sigma_j$ are respectively the numbers of fermions/bosons and qubit excitation at site $j$. Thus, we have
\begin{align}
\label{S7}
\hat{W}|\underline{n\sigma}\rangle&=e^{-i\theta_{\underline{n\sigma}}}|\underline{n\sigma}\rangle,\\
\label{S8}
\theta_{\underline{n\sigma}}&=\sum_{i=1}^L(\omega_i n_i+\Omega_i\sigma_i)+\sum_{i<j}U_{ij}n_in_j.
\end{align}
We derive $\text{tr}\hat{U}^t$ by inserting the identity operator $\sum_{\underline{n\sigma}_{\mu}}|\underline{n\sigma}_{\mu}\rangle\langle\underline{n\sigma}_{\mu}|=\mathds{1}_{\mathcal{N}}$, at different time steps $\mu=1,2,...,t$:
\begin{align}
\label{S9}
\text{tr} \hat{U}^t&=\sum_{\underline{n\sigma}_1,...,\underline{n\sigma}_t}\langle\underline{n\sigma}_1|\hat{V}\hat{W}|\underline{n\sigma}_2\rangle\langle\underline{n\sigma}_2|\hat{V}\hat{W}...|\underline{n\sigma}_t\rangle\notag\\
					&\qquad\qquad\qquad\qquad\times\langle\underline{n\sigma}_t|\hat{V}\hat{W}|\underline{n\sigma}_1\rangle,
\end{align}
where trace requires periodic boundary condition (PBC) in time, $\underline{n\sigma}_{t+1}=\underline{n\sigma}_1$. Using Eq.~\ref{S7} in Eq.~\ref{S9}, we get for the SFF:
\begin{align}
\label{S10}
K(t)&=\sum_{\underline{n\sigma}_1,...,\underline{n\sigma}_t}\sum_{\underline{n\sigma}_1',...,\underline{n\sigma}_t'}\langle e^{-i\sum_{\mu=1}^t\left(\theta_{\underline{n\sigma}_\mu}-\theta_{\underline{n\sigma}_\mu'}\right)}\rangle\notag\\
	&\qquad\times\prod_{\mu=1}^tV_{\underline{n\sigma}_\mu,\underline{n\sigma}_{\mu+1}}V^*_{\underline{n\sigma}_\mu',\underline{n\sigma}_{\mu+1}'}.
\end{align}
For the randomness in the parameters of $\hat{H}_0$ and the long range interaction, we take $\theta_{\underline{n\sigma}_\mu}$(mod $2\pi$) as uniform iid's over $[0,2\pi)$. The last consideration is a random phase approximation (RPA). Within the RPA, we approximate $\langle e^{-i\sum_{\mu=1}^t\left(\theta_{\underline{n\sigma}_\mu}-\theta_{\underline{n\sigma}_\mu'}\right)}\rangle\simeq \delta_{\{\underline{n\sigma}_1,...,\underline{n\sigma}_t\},\{\underline{n\sigma}_1',...,\underline{n\sigma}_t'\}}$, where
\begin{align}
\label{S11}
\delta_{\{\underline{n\sigma}_1,...,\underline{n\sigma}_t\},\{\underline{n\sigma}_1',...,\underline{n\sigma}_t'\}}&=\begin{cases}
1, \text{ if $\{\underline{n\sigma}_1',...,\underline{n\sigma}_t'\}$ is a}\\
\;\;\;\;\;\text{permutation of}\\
\;\;\;\;\;\text{$\{\underline{n\sigma}_1,...,\underline{n\sigma}_t\}$}\\
0, \text{ otherwise}
\end{cases}.
\end{align}
For $t\ll t_\text{H}$ (where $t_\text{H}$ is the Heisenberg time) when a fraction ($\approx t^2/t_\text{H}+\mathcal{O}\left(1/t_\text{H}^2\right)$) of configurations in $\{\underline{n\sigma}_1,...,\underline{n\sigma}_t\}$ with repeated basis vectors is negligible, \textcite{KosPRX2018} have also shown that such configurations with repeated basis do not contribute to leading order term in the SFF. We also perform exact numerical computations of the SFF using Eq.~\ref{S4} to compare it to that obtained using RPA. We show these comparisons in Figs.~\ref{F2},\ref{F5} to validate the RPA analysis for certain parameter regimes of $\hat{H}_0$ in finite systems. Therefore, we can safely assume that the relevant permutations are, $\pi\in S_t$, where, $S_t$, is the permutation group of $t$ distinct objects. Thus, the Eq.~\ref{S10} reduces to 
\begin{align}
\label{S12}
K(t)&=\sum_{\underline{n\sigma}_1,...,\underline{n\sigma}_t}\sum_{\pi\in S_t}\prod_{\mu=1}^tV_{\underline{n\sigma}_\mu,\underline{n\sigma}_{\mu+1}}V^*_{\underline{n\sigma}_{\pi(\mu)},\underline{n\sigma}_{\pi(\mu+1)}}.
\end{align}
Considering $t$ cyclic and $t$ anti-cyclic permutations, the SFF can be written as
\begin{align}
\label{S13}
K(t)=2t\sum_{\underline{n\sigma}_1,...,\underline{n\sigma}_t}\prod_{\mu=1}^t|V_{\underline{n\sigma}_\mu,\underline{n\sigma}_{\mu+1}}|^2=2t\text{ tr}\mathcal{M}^t,
\end{align}
where, the elements of $\mathcal{M}$ are $\mathcal{M}_{\underline{n\sigma},\underline{n\sigma}'}=|\langle\underline{n\sigma}|\hat{V}|\underline{n\sigma}'\rangle|^2$. Here, $\mathcal{M}$ is a doubly stochastic matrix due to unitarity of $\hat{V}$, therefore, the largest eigenvalue of $\mathcal{M}$ is 1 and other eigenvalues ($\lambda_1,...,\lambda_{\mathcal{N}-1}$) satisfy, $1\geq |\lambda_1|\geq |\lambda_2|\geq...\geq|\lambda_{\mathcal{N}-1}|$. Now, the SFF in Eq.~\ref{S13} in terms of eigenvalues of $\mathcal{M}$ reads
\begin{align}
\label{S14}
K(t)=2t\left(1+\sum_{i=1}^{\mathcal{N}-1}\lambda_i^t\right),
\end{align}
which is given in the main paper. For circular orthogonal ensemble (COE), the leading order behavior of universal RMT form $2t$, which appears at long time beyond the Thouless timescale $t^*(L)$ when the second term, $\sum_{i=1}^{\mathcal{N}-1}\lambda_i^t$, diminishes as $|\lambda_i|<1$. Since, $\lambda_i^t$ vanishes faster for smaller value of $|\lambda_i|$, the nature of Thouless time is mainly determined by the largest eigenvalues. To find Thouless time, $t^*(L)$, we study the eigenspectrum of $\mathcal{M}$ in the Trotter regime of small parameters of the driving Hamiltonian. We write $\mathcal{M}$ in terms of the Hadamard product of $\hat{H}_{\text{JC/R}}$ by Taylor expanding $\hat{V}$:
\begin{align}
\label{S15}
\mathcal{M}=\sum_{k=0}^\infty\sum_{l=-k}^k(-1)^l\frac{\left(\hat{H}_{\text{JC/R}}\right)^{k+l}\bullet \left(\hat{H}_{\text{JC/R}}\right)^{k-l}}{(k+l)!(k-l)!}.
\end{align}
We approximate $\mathcal{M}$ by keeping the terms in Eq.~\ref{S15} upto two leading orders ($k=0,1$) in the Trotter regime:
\begin{align}
\label{S16}
\mathcal{M}=\mathds{1}+\hat{H}_{\text{JC/R}}\bullet \hat{H}_{\text{JC/R}}-(\hat{H}_{\text{JC/R}})^2\bullet \mathds{1}+\mathcal{O}(\hat{H}_{\text{JC/R}}^4).
\end{align}
We explicitly calculate $\mathcal{M}$ in the Trotter regime for different types of mixing between fermions or bosons and qubits investigated in the main paper.

\section{Jaynes-Cummings mixing between fermions and qubits}\label{JCFermion}
The driving Hamiltonian $\hat{H}_{\text{JC}}$ for Jaynes-Cummings $(JC)$ mixing between fermions and qubits is given in Eq. (2) of the main paper.
\begin{align}
\label{S17}
\hat{H}_{\text{JC}}=\sum_{i=1}^L\left(g \hat{a}_i^\dagger\hat{\sigma}_i-J\hat{a}_i^\dagger \hat{a}_{i+1}+{\rm H.c.}\right).
\end{align}
We rewrite the fermion creation and annihilation operators in terms of spin-1/2 operators using the Jordan-Wigner transformation.
\begin{align}
\label{S18}
\hat{a}_i=\prod_{j=1}^{i-1}\left(-\hat{\tau}_j^z \right)\hat{\tau}_i\:,~~\hat{a}_i^\dagger =\hat{\tau}_i^\dagger\prod_{j=1}^{i-1}\left(-\hat{\tau}_j^z \right),
\end{align} 
where $\hat{\tau}_j=\left(\hat{\tau}_j^x-i\hat{\tau}_j^y\right)/2$ and $\hat{\tau}_j^\dagger=\left(\hat{\tau}_j^x+i\hat{\tau}_j^y\right)/2$ are the spin-1/2 lowering and raising operators, respectively. Here, $\hat{\tau}_j^{x,y,z}$ are the Pauli matrices at $j$th site. Substituting Eq.~\ref{S18} in Eq.~\ref{S17}, we get with PBC:
\begin{align}
\label{S20}
\hat{H}_{\text{JC}}=g\sum_{i=1}^L\left(\hat{\tau}_i^\dagger\prod_{j=1}^{i-1}\left(-\hat{\tau}_j^z\right)\hat{\sigma}_i+\prod_{j=1}^{i-1}\left(-\hat{\tau}_j^z\right)\hat{\tau}_i\hat{\sigma}_i^\dagger\right)\notag\\
	-J\sum_{i=1}^{L-1}\left(\hat{\tau}_i^\dagger\hat{\tau}_{i+1}+\hat{\tau}_{i+1}^\dagger\hat{\tau}_i\right)+J(-1)^{\hat{N}_f}\left(\hat{\tau}_L^\dagger\hat{\tau}_1+\hat{\tau}_1^\dagger\hat{\tau}_L\right),
\end{align}
where, $\hat{N}_{\text f}=\sum_{i=1}^L\hat{a}_i^\dagger \hat{a}_i=\sum_{i=1}^L\hat{\tau}_i^\dagger\hat{\tau}_i$, is the total number of fermions or corresponding qubit excitations. We notice that the chosen occupation number basis are equivalent to the eigenbasis of $\prod_{i=1}^L\hat{\tau}_i^z\hat{\sigma}_i^z$. To calculate $\hat{H}_{\text{JC}}\bullet \hat{H}_{\text{JC}}$, we need to find the matrix elements of $\hat{H}_{\text{JC}}$. In the chosen basis, $\prod_{j=1}^{i-1}\left(-\hat{\tau}_j^z\right)|\underline{n\sigma}\rangle=\pm|\underline{n\sigma}\rangle$, and the operators $\hat{\tau}_i,\hat{\tau}_i^\dagger,\hat{\sigma}_i,\hat{\sigma}_i^\dagger$ flip the related spin states. Thus, the nonzero matrix elements of  $\hat{\tau}_i^\dagger\prod_{j=1}^{i-1}\left(-\hat{\tau}_j^z\right)\hat{\sigma}_i$ are $\pm1$. Similarly, the nonzero matrix elements of $\hat{\tau}_i^\dagger\hat{\tau}_{i+1}$ are 1. Therefore,
\begin{align}
\label{S21}
(\hat{\tau}_i^\dagger\prod_{j=1}^{i-1}\left(-\hat{\tau}_j^z\right)\hat{\sigma}_i)\bullet(\hat{\tau}_{i'}^\dagger\prod_{j=1}^{i'-1}\left(-\hat{\tau}_j^z\right)\hat{\sigma}_{i'})=\hat{\tau}_i^\dagger\hat{\sigma}_i\delta_{ii'},\\
\label{S22}
(\hat{\tau}_i^\dagger\hat{\tau}_{i+1})\bullet(\hat{\tau}_{i'}^\dagger\hat{\tau}_{i'+1})=\hat{\tau}_i^\dagger\hat{\tau}_{i+1}\delta_{ii'}.
\end{align}
The Eqs.~\ref{S21},\ref{S22} are true in the chosen basis but not in an arbitrary basis. This implies that
\begin{align}
\label{S23}
\hat{H}_{\text{JC}}\bullet \hat{H}_{\text{JC}}=\sum_{i=1}^L(g^2(\hat{\tau}_i^\dagger\hat{\sigma}_i+\hat{\tau}_i\hat{\sigma}_i^\dagger)+J^2(\hat{\tau}_i^\dagger\hat{\tau}_{i+1}+\hat{\tau}_i\hat{\tau}_{i+1}^\dagger))\notag\\
				=\sum_{i=1}^L\left(\frac{g^2}{2}(\hat{\tau}_i^x\hat{\sigma}_i^x+\hat{\tau}_i^y\hat{\sigma}_i^y)+\frac{J^2}{2}(\hat{\tau}_i^x\hat{\tau}_{i+1}^x+\hat{\tau}_i^y\hat{\tau}_{i+1}^y)\right).
\end{align}
\begin{figure*}[t!]
\centering
\subfloat[]{{\includegraphics[scale=0.5]{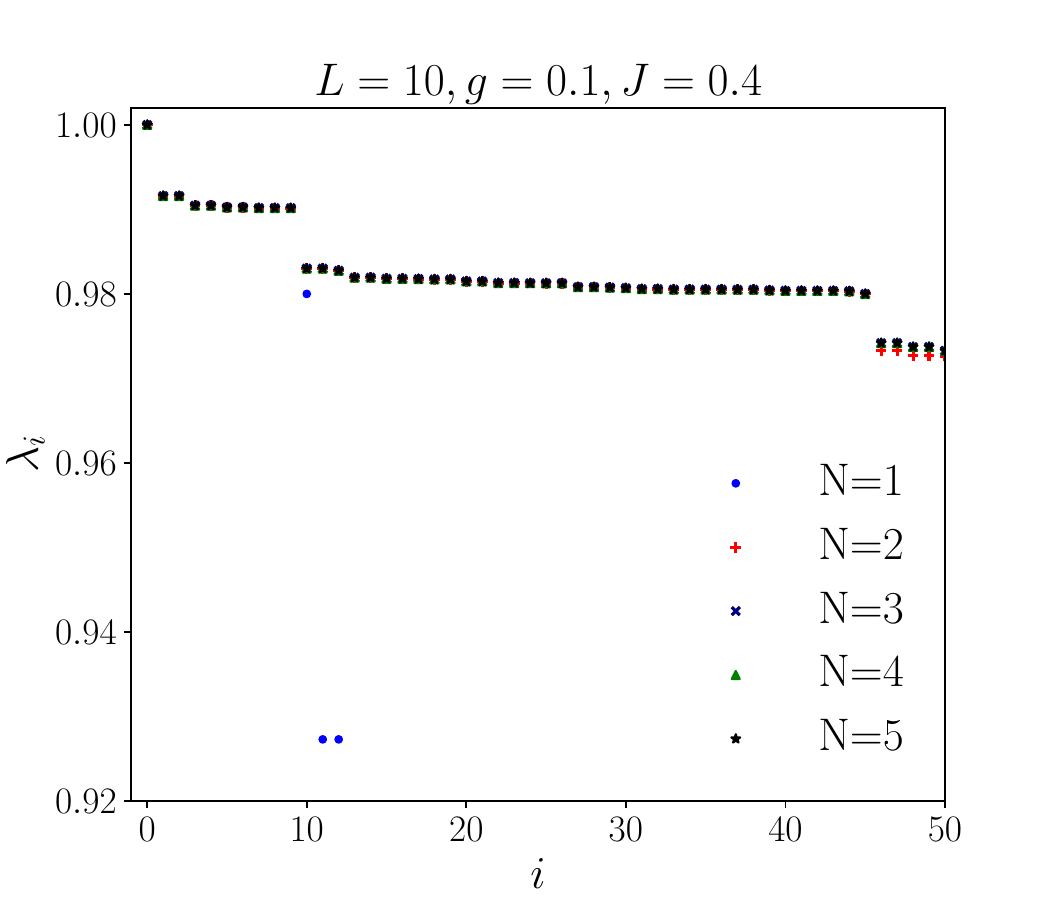}}}
\subfloat[]{{\includegraphics[scale=0.5]{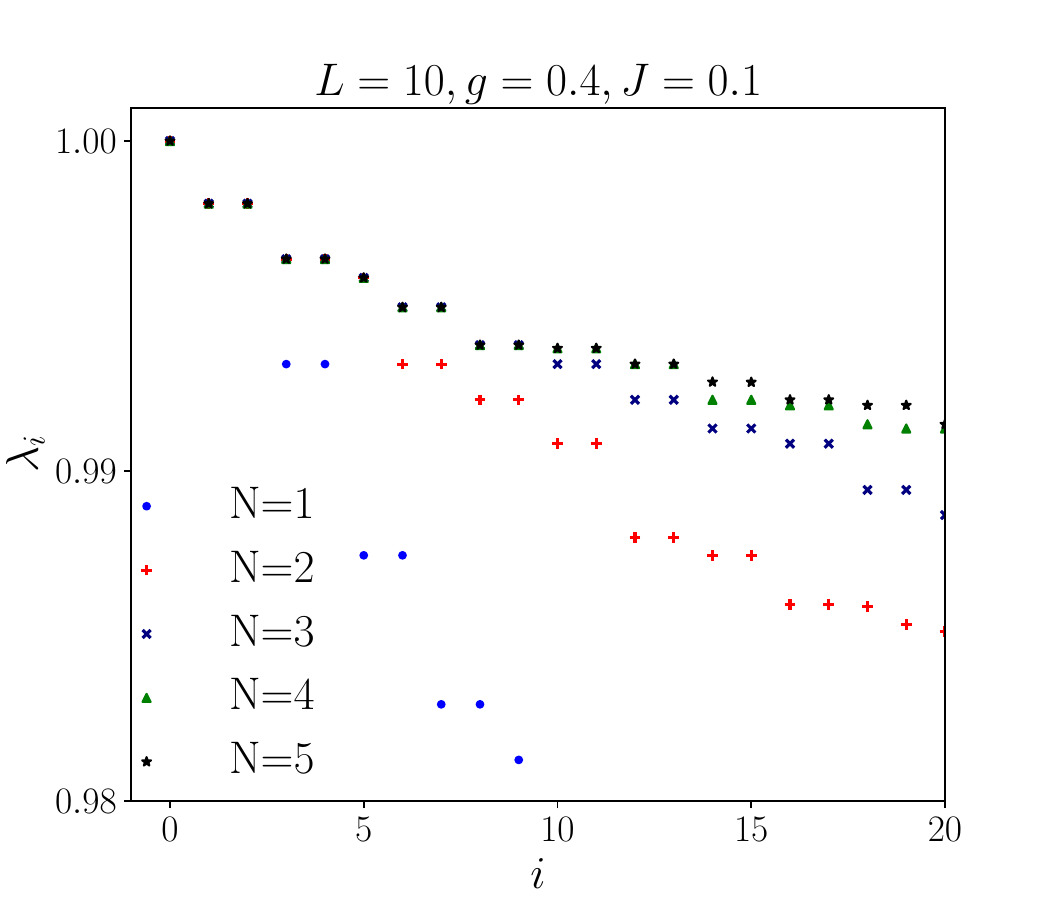}}}
\caption{The largest eigenvalues $\lambda_i$ of $\mathcal{M}_{\text{JC}}^{\text F}$ with $i\in\{0,1,2,...,49\}$ for $g=0.1,J=0.4$ in $(a)$, and $i\in\{0,1,2,...,20\}$ for $g=0.4,J=0.1$ in $(b)$. The plots $(a)$ and $(b)$ show respectively $L$ and $3$ largest $\lambda_i$ being the same for different $N$.}
\label{F1}
\end{figure*}
Since $(\hat{H}_{\text{JC}})^2\bullet \mathds{1}$ is a matrix of diagonal elements of $(\hat{H}_{\text{JC}})^2$, we find
\begin{align}
\label{S24}
(\hat{H}_{\text{JC}})^2\bullet \mathds{1}=\sum_{i=1}^L\left(g^2\frac{1-\hat{\tau}_i^z\hat{\sigma}_i^z}{2}+J^2\frac{1-\hat{\tau}_i^z\hat{\tau}_{i+1}^z}{2}\right).
\end{align}
Plugging Eqs.~\ref{S23},\ref{S24} in Eq.~\ref{S16}, we get
\begin{align}
\label{S25}
\mathcal{M}_{\text{JC}}^{\text F}=&\left(1-\frac{(g^2+J^2)L}{2}\right)\mathds{1}_{\mathcal{N}_{\text{JC}}^{\text F}}\notag\\&+\sum_{i=1}^L\sum_\nu\left(\frac{J^2}{2}\hat{\tau}_i^\nu\hat{\tau}_{i+1}^\nu+\frac{g^2}{2}\hat{\tau}_i^\nu\hat{\sigma}_i^\nu\right)+\mathcal{O}\left(J^4,g^4\right),
\end{align}
where $\nu=x,y,z$ and $\hat{\sigma}_j^x=\hat{\sigma}_j^\dagger+\hat{\sigma}_j,\hat{\sigma}_j^y=\left(\hat{\sigma}_j^\dagger-\hat{\sigma}_j\right)/i$.

The generating Hamiltonian $\mathcal{M}_{\text{JC}}^{\text F}$ commutes with the operators $\sum_i\left(\hat{\tau}_i^\nu+\hat{\sigma}_i^\nu\right)/2$ for $\nu=x,y,z$. These operators satisfy $SU(2)$ algebra, which suggests a $SU(2)$ symmetry of $\mathcal{M}_{\text{JC}}^{\text F}$. The operator for total number of excitations in the present model is $\hat{N}=\sum_{i=1}^L\left(\hat{a}_i^\dagger \hat{a}_i+\hat{\sigma}_i^\dagger\hat{\sigma}_i\right)=\sum_{i=1}^L\left(\hat{\tau}_i^\dagger \hat{\tau}_i+\hat{\sigma}_i^\dagger\hat{\sigma}_i\right)$. The total number of excitations in a state can be changed by the action of operators, $\hat{S}^+=\sum_j\left(\hat{\tau}_j^\dagger+\hat{\sigma}_j^\dagger\right),\hat{S}^-=\sum_j\left(\hat{\tau}_j+\hat{\sigma}_j\right)$. Since,  $ [\hat{N}, \hat{S}^\pm ]=\pm \hat{S}^\pm$, the application of $\hat{S}^\pm$ once on a state leads to a change of the total number of excitations by $\pm1$, respectively. We notice $\hat{S}^\pm$ commutes with $\mathcal{M}_{\text{JC}}^{\text F}$. Thus, if $|\psi\rangle$ is an eigenstate of $\mathcal{M}_{\text{JC}}^{\text F}$ with an eigenvalue $\lambda$ and total excitation $N=1$, $\hat{S}^+|\psi\rangle$ is another eigenstate with the same eigenvalue but $N=2$.  We can then construct eigenstates of $\mathcal{M}_{\text{JC}}^{\text F}$ with a higher number of total excitations but with the same eigenvalue $\lambda$ by repeating the applications of $\hat{S}^+$.

We find from the numerics in Fig.~\ref{F1} that the second largest eigenvalue $\lambda_1$ of $\mathcal{M}_{\text{JC}}^{\text F}$ is the same for $N=1,...,2L-1$. We further notice that $L$ largest eigenvalues of $\mathcal{M}_{\text{JC}}^{\text F}$ are the same for any $N$ in finite-length chains when $g=0.1,J=0.4$. The $L$ largest eigenvalues can be computed analytically for $N=1$ sector for any value of $g,J$. They are
\begin{align}
\label{S26}
\lambda_i&=1-g^2-J^2\left(1-\cos\frac{2\pi i}{L}\right)\notag\\
		&\quad+\sqrt{J^4\left(1-\cos\frac{2\pi i}{L}\right)^2+g^4},
\end{align}
where $i=0,1,...,L-1$. We have $\lambda_0=1$ for $i=0$ as expected for a  double stochastic square matrix. The eigenvalues $\lambda_1,\lambda_2,\dots\lambda_{L-1}$ in Eq.~\ref{S26} are approximately degenerate when $g/J\ll 1$ as shown in Fig.~\ref{F1}(a). We try to find how the Thouless time scales with system size $L$ using these eigenvalues. We approximate the SFF in Eq.~\ref{S14} as
\begin{align}
\label{S27}
K(t)=2t\left(1+\sum_{i=1}^{L-1}\lambda_i^t\right).
\end{align}
We can approximate $\lambda_i$ in Eq.~\ref{S26} for $g/J\ll 1$ and finite $L$ as
\begin{align}
\label{S29}
\lambda_i\approx 1-g^2+\frac{g^4}{2J^2\left(1-\cos\frac{2\pi i}{L}\right)}.
\end{align}
Plugging $\lambda_i$ from Eq.~\ref{S29} in Eq.~\ref{S27}, the SFF reads 
\begin{align}
\label{S30}
K(t)&=2t\left(1+\sum_{i=1}^{L-1}\left(1-g^2+\frac{g^4}{2J^2\left(1-\cos\frac{2\pi i}{L}\right)}\right)^t\right).
\end{align}
We can further simplify the second part of SFF in Eq.~\ref{S30} since $1-g^2\gg \frac{g^4}{2J^2\left(1-\cos\frac{2\pi i}{L}\right)}$ for $g/J\ll 1$ and finite $L$. We thus write
\begin{align}
\label{S32}
&\sum_{i=1}^{L-1}\left(1-g^2+\frac{g^4}{2J^2\left(1-\cos\frac{2\pi i}{L}\right)}\right)^t\notag\\
&=(1-g^2)^{t}(L-1)+\frac{tg^4(1-g^2)^{t-1}}{4J^2}\sum_{i=1}^{L-1}\frac{1}{\sin^2\left(\frac{\pi i}{L}\right)}\notag\\
&=(1-g^2)^{t}(L-1)+\frac{tg^4(1-g^2)^{t-1}}{4J^2}\frac{L^2-1}{3},
\end{align}
where we apply the identity $\sum_{i=1}^{L-1}\csc^2\left(\frac{\pi i}{L}\right)=\frac{L^2-1}{3}$ in the last line. The Thouless time $t^*$ is defined as the time $t$ when the contribution in Eq.~\ref{S32} becomes order of one. Thus, we have 
\begin{align}
\label{S34}
(1-g^2)^{t^*-1}\left((L-1)(1-g^2)+\frac{t^*g^4}{4J^2}\frac{L^2-1}{3}\right)\approx 1.
\end{align}
From the first part of left hand side of Eq.~\ref{S34}, we find $t^*(L)\approx -\log(L-1)/\log(1-g^2)$. To get a better estimate of $L$-dependence of $t^*(L)$ including the second part in Eq.~\ref{S34}, we numerically solve the above equation to find $t^*$ for different $L \in [6,12] $ when $g=0.1,J=0.4$. 
Then we fit the data to find $L$-dependence of $t^*(L)$ to get $t^*(L)=129.24116\log(L+0.10467)-67.74830$, which implies that $t^*(L)\sim \log(L+0.10467)-0.52420$ explaining the $\log(L)$ scaling of $t^*(L)$ observed in our numerical study of the SFF using the eigenvalues of $\mathcal{M}_{\text{JC}}^{\text F}$ in the main paper.

Even for $g/J\ll 1$, the above analysis only works for finite system sizes because of the requirement, $g^4\ll\text{min}\left(J^4\left(1-\cos\frac{2\pi i}{L}\right)\right)$, which breaks down in the thermodynamic limit of $L\rightarrow\infty$ for any finite $i$. For $L\rightarrow\infty$, the second-largest eigenvalue, $\lambda_1\approx 1-\frac{2\pi^2J^2}{L^2}$, determines the Thouless-time scaling for any value of $g,J$ as the largest eigenvalues $\lambda_i$ are no longer nearly degenerate. By relating $\lambda_1^{t^*}\approx 1/e$, we find
\begin{align}
\label{S35}
t^*\approx -1/\log \lambda_1\approx \frac{L^2}{2\pi^2J^2} \propto \mathcal{O}(L^2).
\end{align}
For finite system sizes, the largest eigenvalues $\lambda_i$ are no longer nearly degenerate when $g/J\gg 1$ as shown in Fig.~\ref{F1}(b). Thus, the scaling of $t^*(L)$ with $L$  is again determined by $\lambda_1\approx 1-\frac{2\pi^2J^2}{L^2}$, which again leads to $t^*(L)\propto L^2$. Thus, we find a change in the $L$-dependence of $t^*(L)$ in finite-sized systems when the ratio of  mixing and hopping parameters are increased from low to high value. We discuss such a crossover in Fig.~1 of the main paper.

\begin{figure}[H]
\centering
\includegraphics[width=\linewidth]{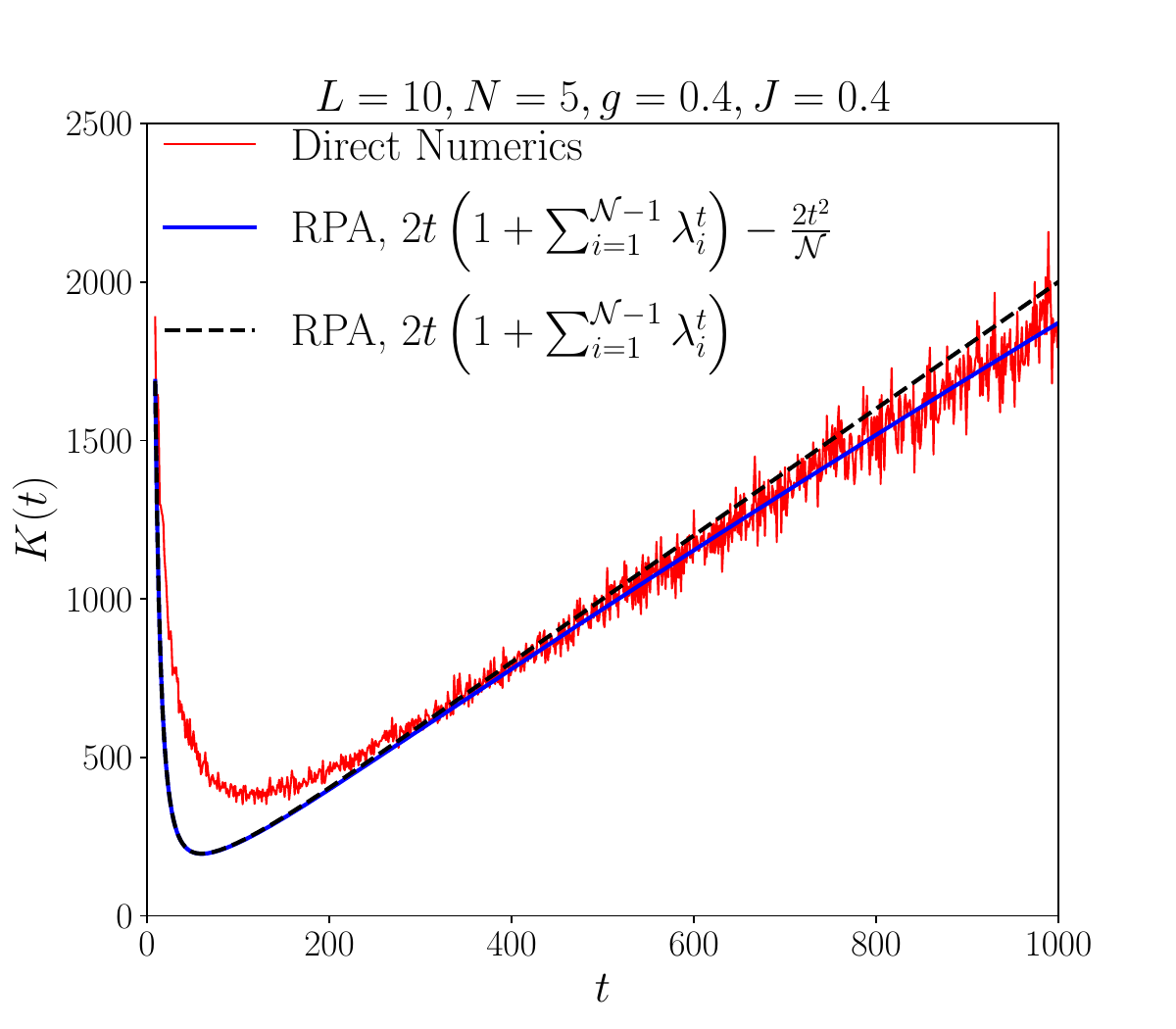}
\caption{Comparison between the exact numerically computed SFF, $K(t)$ vs. $t$, with that obtained using the RPA for Jaynes-Cummings mixing between fermions and qubits. The red curve is exact SFF computed numerically using Eq.~\ref{S4} and the blue (black dashed) curve is that calculated using the first-order and second-order term (only the first-order term) in time within the RPA. All, $\mathcal{N}_{\text{JC}}^{\text F}=15504$, eigenvalues of $\mathcal{M}_{\text{JC}}^{\text F}$ are used for the RPA result. For exact numerical computation, we fix $U_0=10,\alpha=1.4$, and $\omega_i,\Omega_i$ are chosen as Gaussian random variables with a mean $\langle\omega_i\rangle=\langle\Omega_i\rangle=1$ and a standard deviation $\sigma_{\omega_i}=\sigma_{\Omega_i}=0.3$. Averaging over $520$ realizations of disorder is performed for the direct SFF computation.}
\label{F2}
\end{figure}

In the main paper, we show $K(t)$ vs. $t$ calculated using $\lambda_i$ of $\mathcal{M}_{\text{JC}}^{\text F}$ obtained applying the RPA. We numerically compute the SFF for  $JC$ mixing between fermions and qubits using Eq.~\ref{S4} to compare it to that obtained within the RPA using the identity permutation yielding the first-order term in time and also the second-order term of RMT form \cite{KosPRX2018}. We show a good match between two different computations in Fig.~\ref{F2} to validate the RPA analysis in finite-length chains for high values of long-range interaction and random energies to ensure the RPA. We clarify here that we use the full $\mathcal{M}_{\text{JC}}^{\text F}$ instead of its Trotter-regime form in Eq.~\ref{S25} for our numerics in Fig.~\ref{F2}.

\section{Jaynes-Cummings mixing between bosons and qubits}\label{JCBoson}
\begin{figure}
\centering
\subfloat{{\includegraphics[scale=0.5]{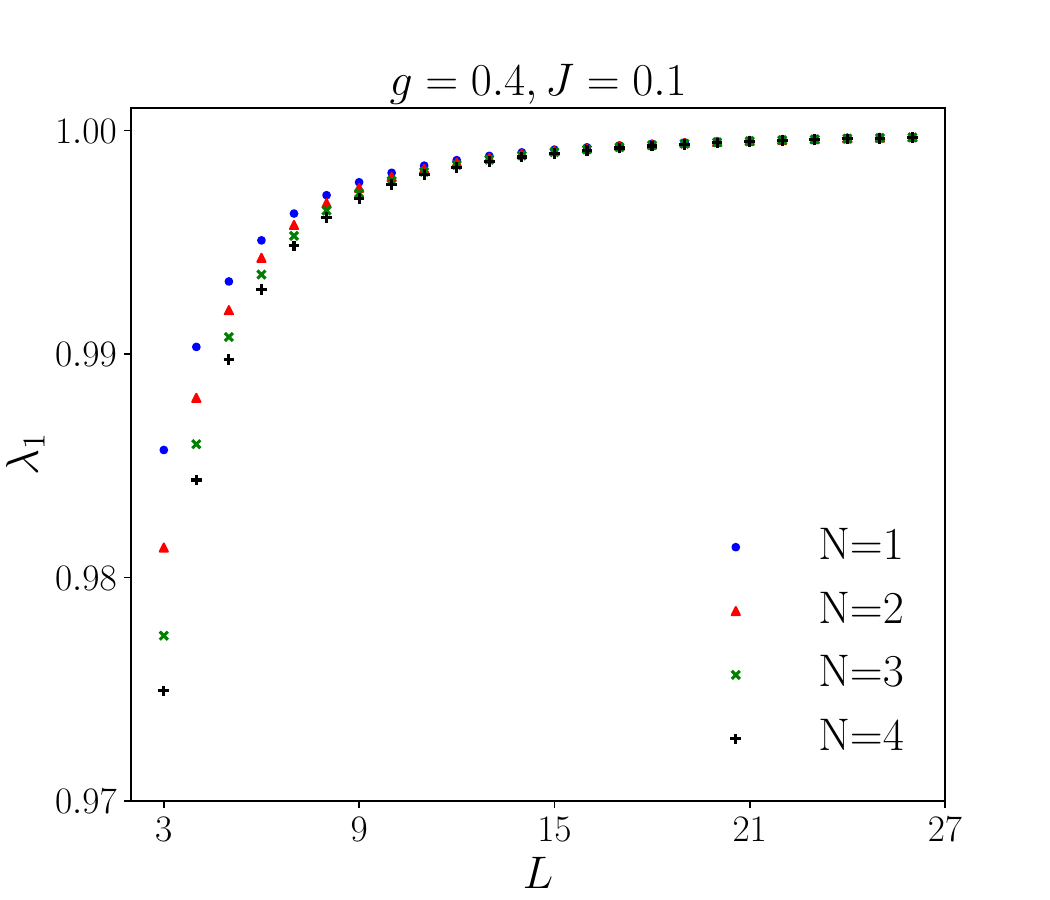}}}
\caption{The second-largest eigenvalue $\lambda_1$ of $\mathcal{M}_{\text{JC}}^{\text{B}}$ vs. $L$ for different total excitations $N$ when $g=0.4,J=0.1$. $\lambda_1$ for different $N$'s approach the same value with an increasing $L$ suggesting emergence of an approximate symmetry of $\mathcal{M}_{\text{JC}}^{\text{B}}$. }
\label{F3}
\end{figure}

The driving Hamiltonian for $JC$ mixing between bosons and qubits is the same as $\hat{H}_{\text{JC}}$ in Eq.~\ref{S17},  where $\hat{a}_i^\dagger,\hat{a}_i$ are now bosonic creation and annihilation operators. Similar to the fermionic case,  the hopping (and/or mixing) terms in $\hat{H}_{\text{JC}}$ at any two different bonds (and/or sites) have no simultaneously non-zero matrix elements in the occupation number basis. Therefore, we get
\begin{align}
\label{S37}
\hat{H}_{\text{JC}}\bullet \hat{H}_{\text{JC}} =g^2\sum_{i=1}^L\left(\hat{a}_i^\dagger \hat{\sigma}_i\bullet \hat{a}_i^\dagger\hat{\sigma}_i+\hat{a}_i\hat{\sigma}_i^\dagger\bullet \hat{a}_i\hat{\sigma}_i^\dagger\right)\notag\\
				+J^2\sum_{i=1}^L\left(\hat{a}_i^\dagger \hat{a}_{i+1}\bullet \hat{a}_i^\dagger \hat{a}_{i+1}+\hat{a}_i \hat{a}_{i+1}^\dagger\bullet \hat{a}_i \hat{a}_{i+1}^\dagger\right).
\end{align}
Since $\langle\underline{n\sigma}|\hat{a}_i^\dagger \hat{\sigma}_i\bullet \hat{a}_i^\dagger\hat{\sigma}_i|\underline{n\sigma}'\rangle=\langle\underline{n\sigma}|\hat{a}_i^\dagger \hat{\sigma}_i|\underline{n\sigma}'\rangle^2=\langle\underline{n\sigma}|\sqrt{\hat{n}_i}\hat{a}_i^\dagger \hat{\sigma}_i|\underline{n\sigma}'\rangle$, and $\langle\underline{n\sigma}|\hat{a}_i^\dagger \hat{a}_{i+1}\bullet \hat{a}_i^\dagger \hat{a}_{i+1}|\underline{n\sigma}'\rangle=\langle\underline{n\sigma}|\hat{a}_i^\dagger \hat{a}_{i+1}|\underline{n\sigma}'\rangle^2=\langle\underline{n\sigma}|\sqrt{\hat{n}_i} \hat{a}_i^\dagger \hat{a}_{i+1}\sqrt{\hat{n}_{i+1}}|\underline{n\sigma}'\rangle$, we can simplify as
\begin{align}
\label{S38}
\hat{H}_{\text{JC}}\bullet \hat{H}_{\text{JC}} =g^2\sum_{i=1}^L(\sqrt{\hat{n}_i}\hat{a}_i^\dagger\hat{\sigma}_i+\hat{a}_i\sqrt{\hat{n}_i}\hat{\sigma}_i^\dagger)\notag\\
				+J^2\sum_{i=1}^L(\sqrt{\hat{n}_i}\hat{a}_i^\dagger \hat{a}_{i+1}\sqrt{\hat{n}_{i+1}}+\sqrt{\hat{n}_{i+1}}\hat{a}_{i+1}^\dagger \hat{a}_i\sqrt{\hat{n}_i}).
\end{align}
Since $(\hat{H}_{\text{JC}})^2\bullet \mathds{1}$ is a matrix of diagonal elements of $(\hat{H}_{\text{JC}})^2$, we get
\begin{align}
\label{S39}
(\hat{H}_{\text{JC}})^2\bullet \mathds{1}&=g^2\sum_{i=1}^L(\hat{n}_i+\hat{\sigma}_i^\dagger\hat{\sigma}_i)\notag\\
			&+J^2\sum_{i=1}^L\left(\hat{n}_i+\hat{n}_{i+1}+2\hat{n}_i \hat{n}_{i+1}\right).
\end{align}
Substituting Eqs.~\ref{S38},\ref{S39} in Eq.~\ref{S16}, we find
\begin{align}
\label{S40}
\mathcal{M}_{\text{JC}}^{\text{B}}&=\mathds{1}_{\mathcal{N}_{\text{JC}}^{\text{B}}}+g^2\sum_{i=1}^L\left(\sqrt{\hat{n}_i}\hat{a}_i^\dagger\hat{\sigma}_i+\hat{a}_i\sqrt{\hat{n}_i}\hat{\sigma}_i^\dagger-\hat{n}_i-\hat{\sigma}_i\hat{\sigma}_i^\dagger\right)\notag\\
			&+J^2\sum_{i=1}^L\Big(\sqrt{\hat{n}_i}\hat{a}_i^\dagger \hat{a}_{i+1}\sqrt{\hat{n}_{i+1}}+\sqrt{\hat{n}_{i+1}}\hat{a}_{i+1}^\dagger \hat{a}_i\sqrt{\hat{n}_i}\notag\\
			&-\hat{n}_i-\hat{n}_{i+1}-2\hat{n}_i\hat{n}_{i+1}\Big)+\mathcal{O}(J^4,g^4).
\end{align}
The above expression can be written in terms of operators, $\hat{K}_i^-=\sqrt{\hat{n}_i}\hat{a}_i^\dagger,\hat{K}_i^+=\hat{a}_i\sqrt{\hat{n}_i}$, and $\hat{K}_i^0=-\left(\hat{n}_i+1/2\right)$, satisfying $SU(1,1)$ algebra, $[\hat{K}_i^+,\hat{K}_j^-]=-\hat{K}_i^0\delta_{ij},[\hat{K}_i^0,\hat{K}_i^\pm]=\pm \hat{K}_i^\pm\delta_{ij}$, as
\begin{align}
\label{S41}
\mathcal{M}_{\text{JC}}^{\text{B}}&=(1+\frac{(g^2+J^2)L}{2})\mathds{1}_{\mathcal{N}_{\text{JC}}^{\text{B}}}+g^2\sum_{i=1}^L\big(\hat{K}_i^-\hat{\sigma}_i+\hat{K}_i^+\hat{\sigma}_i^\dagger\notag\\&+\hat{K}_i^0-\hat{\sigma}_i^\dagger\hat{\sigma}_i\big)+J^2\sum_{i=1}^L\big(\hat{K}_i^+ \hat{K}_{i+1}^-+\hat{K}_i^- \hat{K}_{i+1}^+\notag\\&-2\hat{K}_i^0 \hat{K}_{i+1}^0\big)+\mathcal{O}(J^4,g^4).
\end{align}
We can further define these operators, $\hat{K}_j^1=(\hat{K}_j^++\hat{K}_j^-)/2,\hat{K}_j^2=(\hat{K}_j^+-\hat{K}_j^-)/2i$, and rewrite Eq.~\ref{S41} as
\begin{align}
\label{S42}
\mathcal{M}_{\text{JC}}^{\text{B}}&=(1+\frac{(g^2+J^2)L}{2})\mathds{1}_{\mathcal{N}_{\text{JC}}^{\text{B}}}+g^2\sum_{i=1}^L\big(\hat{K}_i^1\hat{\sigma}_i^x-\hat{K}_i^2\hat{\sigma}_i^y\notag\\&+\hat{K}_i^0-\hat{\sigma}_i^\dagger\hat{\sigma}_i\big)+2J^2\sum_{i=1}^L\big(\hat{K}_i^1 \hat{K}_{i+1}^1+\hat{K}_i^2 \hat{K}_{i+1}^2\notag\\&-\hat{K}_i^0 \hat{K}_{i+1}^0\big)+\mathcal{O}(J^4,g^4),
\end{align}
where $\hat{\sigma}_i^x=\hat{\sigma}_i^\dagger+\hat{\sigma}_i,\hat{\sigma}_i^y=\left(\hat{\sigma}_i^\dagger-\hat{\sigma}_i\right)/i$. The generating Hamiltonian $\mathcal{M}_{\text{JC}}^{\text{B}}$ commutes with $\sum_{i=1}^L\hat{K}_i^0+\hat{\sigma}_i^\dagger\hat{\sigma}_i$, which suggests a $U(1)$ symmetry of $\mathcal{M}_{\text{JC}}^{\text{B}}$. But unlike the fermionic case, the lowering and raising operators like $\sum_{i=1}^L(\hat{K}_i^++\hat{\sigma}_i),\sum_{i=1}^L(\hat{K}_i^-+\hat{\sigma}_i^\dagger)$ do not commute with $\mathcal{M}_{\text{JC}}^{\text{B}}$, which indicates an absence of $SU(2)$ or $SU(1,1)$ symmetry of $\mathcal{M}_{\text{JC}}^{\text{B}}$. Thus, the second-largest eigenvalue $\lambda_1$ of $\mathcal{M}_{\text{JC}}^{\text{B}}$ is not required to be the same for different total number of excitations $N$. Numerical study in Fig.~\ref{F3} confirms that $\lambda_1$ is different for different $N$ at a fixed $L$ when $g=0.4,J=0.1$. However, Fig.~\ref{F3} also shows that $\lambda_1$ for different $N$ seems to converge with an increasing $L$.  For system sizes $L \in [10,26]$, we numerically find $\lambda_1(N) \approx 1-c_N/L^{2}$ for different $N$'s, and the difference $(\lambda_1(N)-\lambda_1(N+1))$ falls as $c'_N/L^3$, where $c_N$ and $c'_N$ are $N$-dependent constants. These scaling suggests that $\lambda_1$ becomes independent of $N$ at large $L$ due to an emergent approximate symmetry of $\mathcal{M}_{\text{JC}}^{\text{B}}$. A similar $N$-independence of $L$ largest eigenvalues of $\mathcal{M}_{\text{JC}}^{\text{B}}$ is also observed with increasing $L$ when $g/J\ll 1$. 

\begin{figure*}
\centering
\includegraphics[width=\linewidth]{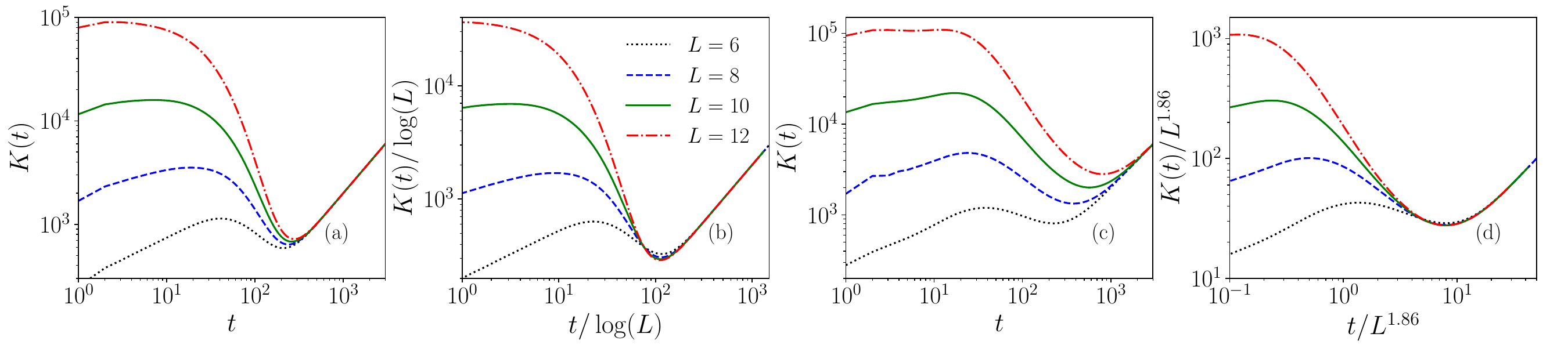}
\caption{Spectral form factor $K(t)$ using Eq.~\ref{S14} for different system sizes $L$ of the kicked chain with $JC$ mixing between bosons and qubits for $g=0.1,J=0.4$ in $(a,b)$, and $g=0.4,J=0.1$ in $(c,d)$. We take half-filling $N/L=1/2$. In $(b)$ and $(d)$, we show data collapse in scaled time $t/\log L$ and $t/L^{1.86}$, respectively.}
\label{F4}
\end{figure*}

Similar to the fermionic case in the earlier section,  $\lambda_i$ of $\mathcal{M}_{\text{JC}}^{\text{B}}$ can be calculated analytically for $N=1$, and $\lambda_i$ are then identical to those in Eq.~\ref{S26}. Incorporating the above observations for an emergent approximate symmetry of $\mathcal{M}_{\text{JC}}^{\text{B}}$ with $\lambda_i$ for $N=1$, we then expect a change in the $L$-dependence of $t^*(L)$ from $\log L$ to $L^2$ with an increasing $g/J$ from much lower than one to much larger than one at any filling fraction $N/L$ in finite-size systems of bosons and qubits. We demonstrate such a crossover in the $L$-scaling of $t^*(L)$ by studying $K(t)$ vs. $t$ within the RPA in Fig.~\ref{F4} at half filling $N/L=1/2$ for finite system sizes.

\section{Rabi mixing between fermions and qubits}
\label{RabiFermion}
The driving Hamiltonian $\hat{H}_{\text{R}}$ for Rabi $(R)$ mixing between fermions and qubits is given in Eq. (3) of the main paper.
\begin{align}
\label{S43}
\hat{H}_{\text{R}}&=g\sum_{i=1}^L (\hat{a}_i^\dagger +\hat{a}_i)(\hat{\sigma}_i^\dagger+\hat{\sigma}_i)-J\sum_{i=1}^L\left(\hat{a}_i^\dagger \hat{a}_{i+1}+{\text{H.c.}}\right).
\end{align}
Following the steps presented for deriving the generating Hamiltonian in the Trotter regime for $JC$ mixing between fermions and qubits in Sec.~\ref{JCFermion}, we  find
\begin{align}
\label{S44}
\mathcal{M}_{\text R}^{\text F}&=\left(1-\frac{(2g^2+J^2)L}{2}\right)\mathds{1}_{\mathcal{N}_{\text R}^{\text F}}+\sum_{i=1}^L\Big(\frac{J^2}{2}\sum_{\nu}\hat{\tau}_i^\nu\hat{\tau}_{i+1}^\nu\notag\\
				&\qquad+g^2\hat{\tau}_i^x\hat{\sigma}_i^x\Big)+\mathcal{O}(J^4,g^4),
\end{align}
where $\nu=x,y,z$. By performing a rotation $\big(\mathcal{R}^y_{\theta} \equiv \exp(i\theta\sum_{j=1}^{L}(\hat{\tau}_j^y+\hat{\sigma}_j^y)/2)\big)$ around $y$-axis by $\theta=\pi/2$, we transform the operators as $\mathcal{R}^y_{\theta}\hat{\tau}_i^x\mathcal{R}_{\theta}^{y\dagger}=\hat{\tilde{\tau}}_i^z,\mathcal{R}^y_{\theta}\hat{\tau}_i^y\mathcal{R}_{\theta}^{y\dagger}=\hat{\tilde{\tau}}_i^y,\mathcal{R}^y_{\theta}\hat{\tau}_i^z\mathcal{R}_{\theta}^{y\dagger}=-\hat{\tilde{\tau}}_i^x,\mathcal{R}^y_{\theta}\hat{\sigma}_i^x\mathcal{R}_{\theta}^{y\dagger}=\hat{\tilde{\sigma}}_i^z,\mathcal{R}^y_{\theta}\hat{\sigma}_i^y\mathcal{R}_{\theta}^{y\dagger}=\hat{\tilde{\sigma}}_i^y,\mathcal{R}^y_{\theta}\hat{\sigma}_i^z\mathcal{R}_{\theta}^{y\dagger}=-\hat{\tilde{\sigma}}_i^x$, which lead to $\tilde{\mathcal{M}}_{\text R}^{\text F}=\mathcal{R}^y_{\theta}\mathcal{M}_{\text R}^{\text F}\mathcal{R}_{\theta}^{y\dagger}$:
\begin{align}
\label{S45}
\tilde{\mathcal{M}}_{\text R}^{\text F}&=\left(1-\frac{(2g^2+J^2)L}{2}\right)\mathds{1}_{\mathcal{N}_{\text R}^{\text F}}+\sum_{i=1}^L\Big(\frac{J^2}{2}\sum_{\nu}\hat{\tilde{\tau}}_i^\nu\hat{\tilde{\tau}}_{i+1}^\nu\notag\\
				&\qquad+g^2\hat{\tilde{\tau}}_i^z\hat{\tilde{\sigma}}_i^z\Big)+\mathcal{O}(J^4,g^4).
\end{align}
The generating Hamiltonian $\tilde{\mathcal{M}}_{\text R}^{\text F}$ commutes with $\hat{\tilde{\sigma}}_i^z$ for each $i\in[1,L]$. We call this symmetry collectively as $u^{\otimes L}(1)$. $\tilde{\mathcal{M}}_{\text R}^{\text F}$ also commutes with $\sum_{i=1}^L\hat{\tilde{\tau}}_i^z$, which is a global $U(1)$ symmetry. So, $\tilde{\mathcal{M}}_{\text R}^{\text F}$ has $U(1)\otimes u^{\otimes L}(1)$ symmetry. Thus, the eigenvalues and eigenstates of $\tilde{\mathcal{M}}_{\text R}^{\text F}$ can be labelled by excitation of individual $\tilde{\sigma}$ qubit, $\tilde{\sigma}_i\left(\equiv \hat{\tilde{\sigma}}_i^\dagger\hat{\tilde{\sigma}}_i\right)$, and total  excitations of $\tilde{\tau}$ qubits, $\tilde{N}_{\text f}\left(\equiv\sum_{i=1}^L\hat{\tilde{\tau}}_i^\dagger\hat{\tilde{\tau}}_i\right)$. The states with the largest eigenvalue $\lambda_0=1$ have $\tilde{N}_{\text f}=0,\tilde{\sigma}_i=0$ or $\tilde{N}_{\text f}=L,\tilde{\sigma}_i=1,$ for all $i\in[1,L]$. The degeneracy of $2$ for the largest eigenvalue is due to the invariance of $\tilde{\mathcal{M}}_{\text R}^{\text F}$ under the transformation, $\prod_{i=1}^L\hat{\tilde{\tau}}_i^x\hat{\tilde{\sigma}}_i^x$. The last operator flips all the spins to transform $\tilde{N}_{\text f}=0\rightarrow L,\tilde{\sigma}_i=0\rightarrow 1$ for all $i\in[1,L]$ leading to the degeneracy of $2$. This is equivalent to $\mathds{Z}_2$ symmetry also present in $\hat{H}(t)$ (the $U(1)$ breaking $\hat{H}_{\text R}$ does not mix between the even and odd number of total excitation sectors). To study chaos, we choose either the even or odd sector of the Hilbert space of $\hat{H}(t)$ only. Similarly, we choose only one largest eigenvalue of $\tilde{\mathcal{M}}_{\text R}^{\text F}$, and study excitations (second-largest eigenvalues) around it.  We here choose the largest eigenvalue state with $\tilde{N}_{\text f}=0,\tilde{\sigma}_i=0,$ for all $i\in[1,L]$. The numerical study for different values of $g,J$ shows that there are two type of $\tilde{\tau}$ and $\tilde{\sigma}$ spin configurations giving the second-largest eigenvalue of $\tilde{\mathcal{M}}_{\text R}^{\text F}$ depending on the ratio $g/J$ in the Trotter regime. We describe them below.

\textbf{Case 1:} For $g/J \ll 1$, the configurations leading to the second-largest eigenvalues appear by one change in the excitation from the largest eigenvalue state, e.g., $\tilde{N}_{\text f}=0,\tilde{\sigma}_j=1,\tilde{\sigma}_{i\neq j}=0,i=1,...,j-1,j+1,...,L$, and $\tilde{N}_{\text f}=1,\tilde{\sigma}_i=0$ for all $i\in[1,L]$. We find
\begin{align}
\label{S46}
&\sum_{i=1}^L\sum_{\nu}\hat{\tilde{\tau}}_i^\nu\hat{\tilde{\tau}}_{i+1}^\nu|\tilde{N}_{\text f}=0,\{\tilde{\sigma}\}=\{0,...,0,1,0,...,0\}\rangle\notag\\
&\qquad\quad=L|\tilde{N}_{\text f}=0,\{\tilde{\sigma}\}=\{0,...,0,1,0,...,0\}\rangle,\\
\label{S47}
&\sum_{i=1}^L\hat{\tilde{\tau}}_i^z\hat{\tilde{\sigma}}_i^z|\tilde{N}_{\text f}=0,\{\tilde{\sigma}\}=\{0,...,0,1,0,...,0\}\rangle\notag\\
&\qquad=(L-2)|\tilde{N}_{\text f}=0,\{\tilde{\sigma}\}=\{0,...,0,1,0,...,0\}\rangle.
\end{align}
Therefore,
\begin{align}
\label{S48}
&\tilde{\mathcal{M}}_{\text R}^{\text F}|\tilde{N}_{\text f}=0,\{\tilde{\sigma}\}=\{0,...,0,1,0,...,0\}\rangle\notag\\
&\qquad=(1-2g^2)|\tilde{N}_{\text f}=0,\{\tilde{\sigma}\}=\{0,...,0,1,0,...,0\}\rangle,
\end{align}
and we get $\lambda_1=1-2g^2$ with $L$-fold degeneracy, which is due to appearance of $\tilde{\sigma}_j=1$ at any $j=1,2,..,L$. For $\tilde{N}_{\text f}=1,\tilde{\sigma}_i=0$, $\tilde{\mathcal{M}}_{\text R}^{\text F}$ becomes the $XXX$ Heisenberg spin-1/2 chain for $\tilde{\tau}$ spin in a uniform external magnetic field ($2g^2$).
\begin{align}
\label{S49}
\tilde{\mathcal{M}}_{\text R}^{\text F}|_{\tilde{N}_{\text f}=1,\tilde{\sigma}_i=0}=\left(1-\frac{(2g^2+J^2)L}{2}\right)\mathds{1}_{\mathcal{N}_{\text R}^{\text F}}\notag\\+\sum_{i=1}^L\left(\frac{J^2}{2}\sum_{\nu}\hat{\tilde{\tau}}_i^\nu\hat{\tilde{\tau}}_{i+1}^\nu-g^2\hat{\tilde{\tau}}_i^z\right)+\mathcal{O}(J^4,g^4).
\end{align}
For total number of $\tilde{\tau}$ excitations, $\tilde{N}_{\text f}=1$, the eigenvalues of $\tilde{\mathcal{M}}_{\text R}^{\text F}$ are
\begin{align}
\label{S50}
\lambda_{k_m}=1-2g^2-2J^2(1-\cos k_m),
\end{align}
where $k_m=2\pi m/L, m=0,1,...,L-1$. For $k_m=0$, we get the second-largest eigenvalue $\lambda_1=1-2g^2$. Therefore, we get total $(L+1)$-fold degeneracy in this case. The Thouless-time scaling in the thermodynamic limit of $L\to \infty$ can be obtained by setting $(L+1)\lambda_1^{t^*}\approx 1$ as
\begin{align}
\label{S51}
t^*(L)\approx -\frac{\log(L+1)}{\log(1-2g^2)} \approx \mathcal{O}(\log L).
\end{align} 
\textbf{Case 2:} For $g/J\gg1$, the configurations leading to the second-largest eigenvalues appear by one change in the excitation from the largest eigenvalue state, e.g., $\tilde{N}_{\text f}=1,\tilde{\sigma}_j=1,\tilde{\sigma}_{i\neq j}=0,i=1,...,j-1,j+1,...,L$. We can cast $\tilde{\mathcal{M}}_{\text R}^{\text F}$ in this case as a tight-binding chain with an impurity in the onsite energy:
\begin{align}
\label{S53}
\tilde{\mathcal{M}}_{\text R}^{\text F}|_{\tilde{N}_{\text f}=1,\tilde{\sigma}_j=1}=1-4g^2-2J^2+4g^2\hat{\tilde{\tau}}_j^\dagger\hat{\tilde{\tau}}_j\notag\\
				+J^2\sum_{i=1}^L(\hat{\tilde{\tau}}_i^\dagger \hat{\tilde{\tau}}_{i+1}+\hat{\tilde{\tau}}_i \hat{\tilde{\tau}}_{i+1}^\dagger).
\end{align}
The spectrum of the Hamiltonian in Eq.~\ref{S53} can be derived using the Dyson equation to treat the impurity as a perturbation following \textcite{economou2006green}. We separate $\tilde{\mathcal{M}}_{\text R}^{\text F}|_{\tilde{N}_{\text f}=1,\tilde{\sigma}_j=1}$ in two parts as 
\begin{align}
\label{S54}
\tilde{\mathcal{M}}_{\text R}^{\text F}|_{\tilde{N}_{\text f}=1,\tilde{\sigma}_j=1}=\mathcal{M}_0+\mathcal{M}_1,
\end{align}
where $\mathcal{M}_0=1-4g^2-2J^2+J^2\sum_{i=1}^L(\hat{\tilde{\tau}}_i^\dagger \hat{\tilde{\tau}}_{i+1}+\hat{\tilde{\tau}}_i \hat{\tilde{\tau}}_{i+1}^\dagger),\mathcal{M}_1=4g^2\hat{\tilde{\tau}}_j^\dagger\hat{\tilde{\tau}}_j=4g^2|j\rangle\langle j|$. Here, $|j\rangle=\hat{\tilde{\tau}}_j^{\dagger}|\varphi\rangle$, where $|\varphi\rangle$ is a vacuum state. 
We consider $G_0(z)$ as the Green's function for $\mathcal{M}_0$, and $G(z)$ as that  for $\mathcal{M}_0+\mathcal{M}_1$. The Dyson equation \cite{economou2006green} gives,
\begin{align}
\label{S55}
G&=G_0\notag\\&~+G_0\left(\mathcal{M}_1+\mathcal{M}_1 G_0 \mathcal{M}_1+\mathcal{M}_1 G_0 \mathcal{M}_1 G_0 \mathcal{M}_1+... \right) G_0\notag\\
	&=G_0+G_0\Big((2g)^2+(2g)^4G_0(j,j;z)\notag\\&\quad+(2g)^6\left(G_0(j,j;z)\right)^2 +... \Big)|j\rangle\langle j| G_0\notag\\
	&=G_0+G_0\frac{(2g)^2}{1-(2g)^2G_0(j,j;z)}|j\rangle\langle j|G_0,
\end{align}
where $G_0(j,j;z)=\langle j|G_0(z)|j\rangle$. We find from Eq.~\ref{S55} that $G$ has an isolated pole at 
\begin{align}
\label{S56}
G_0(j,j;E_b)=\frac{1}{4g^2},
\end{align}
which gives the eigenvalue of the corresponding bound state. The free Green's function can be determined using the eigenvalues and eigenstates of $\mathcal{M}_0$ as $G_0(z)=\sum_{m}|k_m\rangle\langle k_m|/(z-E(k_m))$, where $|k_m\rangle$ are eigenstates of $\mathcal{M}_0$ with eigenvalues $E(k_m)=1-4g^2-2J^2+2J^2\cos k_m,\;k_m=2\pi m/L$ for $m=0,1,2,...,L-1$. Therefore, we get
\begin{align}
\label{S57}
G_0(j_1,j_2;z)=\sum_{m}\frac{\langle j_1|k_m\rangle\langle k_m|j_2\rangle}{z-E(k_m)}=\frac{1}{L}\sum_{m}\frac{e^{ik_m(j_1-j_2)}}{z-E(k_m)}.
\end{align}
It is easy to evaluate the above sum in the thermodynamic limit of $L\to \infty$ as
\begin{align}
\label{S58}
G_0(j_1,j_2;z)=\frac{1}{2\pi}\int_0^{2\pi}dk\frac{e^{ik(j_1-j_2)}}{z-E(k)}\notag\\
				=\frac{1}{2\pi}\int_0^{2\pi}dk\frac{e^{ik(j_1-j_2)}}{z-1+4g^2+2J^2-2J^2\cos k}.
\end{align}
We observe $G_0(j_1,j_2;z)= G_0(j_2,j_1;z)$ as $k\rightarrow 2\pi-k$, which implies that $G_0(j_1,j_2;z)$ is a function of $|j_1-j_2|$. We define $w=e^{ik},\epsilon=(z-1+4g^2+2J^2)/(2J^2)$ to rewrite as
\begin{align}
\label{S59}
G_0(j_1,j_2;z)=-\frac{1}{2\pi iJ^2}\oint dw \frac{w^{|j_1-j_2|}}{w^2-2\epsilon w+1},
\end{align}
where the contour is a unit circle on complex $w$ plane. The integrand has poles at $w=w_\pm=\epsilon\pm\sqrt{\epsilon^2-1}$. We further notice $w_+ w_-=1$, which implies that $|w_+||w_-|=1$. Thus, we get 
\begin{align}
\label{S60}
|w_+|=\begin{cases}
<1,\quad \frac{\pi}{2}<\arg(\epsilon)<\frac{3\pi}{2}\\
>1,\quad \text{otherwise}
\end{cases},\\
\label{S61}
|w_-|=\begin{cases}
>1,\quad \frac{\pi}{2}<\arg(\epsilon)<\frac{3\pi}{2}\\
<1,\quad \text{otherwise}
\end{cases}.
\end{align}
For $\epsilon$ real, and $-1<\epsilon<1$, $|w_\pm|=1$, and the poles exist on the integration contour. This happens for $1-4g^2-4J^2<z<1-4g^2$, which is a range of continuous spectrum of $\mathcal{M}_0$. We have $w_+$ or $w_-$ inside the integration contour depending on the argument of $\epsilon$. We finally get
\begin{align}
\label{S62}
G_0(j_1,j_2;z)=-\frac{1}{J^2}\frac{w_\pm^{|j_1-j_2|}}{w_\pm-w_\mp}.
\end{align} 
Equating Eqs.~\ref{S62} and \ref{S56} at $z=E_b$, we find
\begin{align}
\label{S63}
\epsilon&=\pm\sqrt{1+\frac{4g^4}{J^4}},\\
\label{S64}
\implies E_{b\pm}&=1-4g^2-2J^2\pm 2J^2\sqrt{1+\frac{4g^4}{J^4}}.
\end{align}
We have two bound states in Eq.~\ref{S64}, and the one corresponding to the second-largest eigenvalue of $\tilde{\mathcal{M}}_{\text R}^{\text F}$ in this case is
\begin{align}
\label{S65}
\lambda_1\equiv E_{b+}=1-4g^2-2J^2+ 2J^2\sqrt{1+\frac{4g^4}{J^4}}.
\end{align}
Here, $\tilde{\sigma}_j=1$ can be possible at any one site of $L$ sites, which leads to an $L$-fold degeneracy. Thus, we again get, $t^*(L) \propto \mathcal{O}(\log L)$ for this case also.

There is a transition in the gap between $\lambda_0$ and $\lambda_1$ from \textbf{case 1} to \textbf{case 2} as $g/J$ changes. We can find the precise transition point by equating the eigenvalue obtained for the two cases. The transition point turns out to be
\begin{align}
\label{S66}
\left(\frac{g}{J}\right)_{\text{c}}=\sqrt{\frac{2}{3}}.
\end{align}
We notice that the $L$-fold degeneracy is due to the spatially uniform coupling $(g)$ between fermions and qubits. Instead, we can consider the driven Hamiltonian with site-dependent  coupling $g_i$ at site $i$ for $R$ mixing as
\begin{align}
\label{S67}
\hat{H}_{\text R}&=\sum_{i=1}^L g_i(\hat{a}_i^\dagger +\hat{a}_i)(\hat{\sigma}_i^\dagger+\hat{\sigma}_i)-J\sum_{i=1}^L\left(\hat{a}_i^\dagger \hat{a}_{i+1}+{\text{H.c}}.\right),
\end{align}
which would transform the generating Hamiltonian in Eq.~\ref{S44} as
\begin{align}
\label{S68}
\mathcal{M}_{\text R}^{\text F}&=\big(1-\frac{J^2L}{2}-\sum_{i=1}^Lg_i^2\big)\mathds{1}_{\mathcal{N}_{\text R}^{\text F}}+\sum_{i=1}^L\Big(\frac{J^2}{2}\sum_{\nu}\hat{\tau}_i^\nu\hat{\tau}_{i+1}^\nu\notag\\
				&\qquad+g_i^2\hat{\tau}_i^x\hat{\sigma}_i^x\Big)+\mathcal{O}(J^4,g_i^4).
\end{align}
The second-largest eigenvalues of $\mathcal{M}_{\text R}^{\text F}$ are $1-2g_i^2$. Since $g_i$ is different for different $i$, the $L$-fold degeneracy of the second-largest eigenvalue is lifted for site-dependent $R$ mixing, and we get $t^*(L)\sim L^0$.

\begin{figure}[H]
\centering
\includegraphics[width=\linewidth]{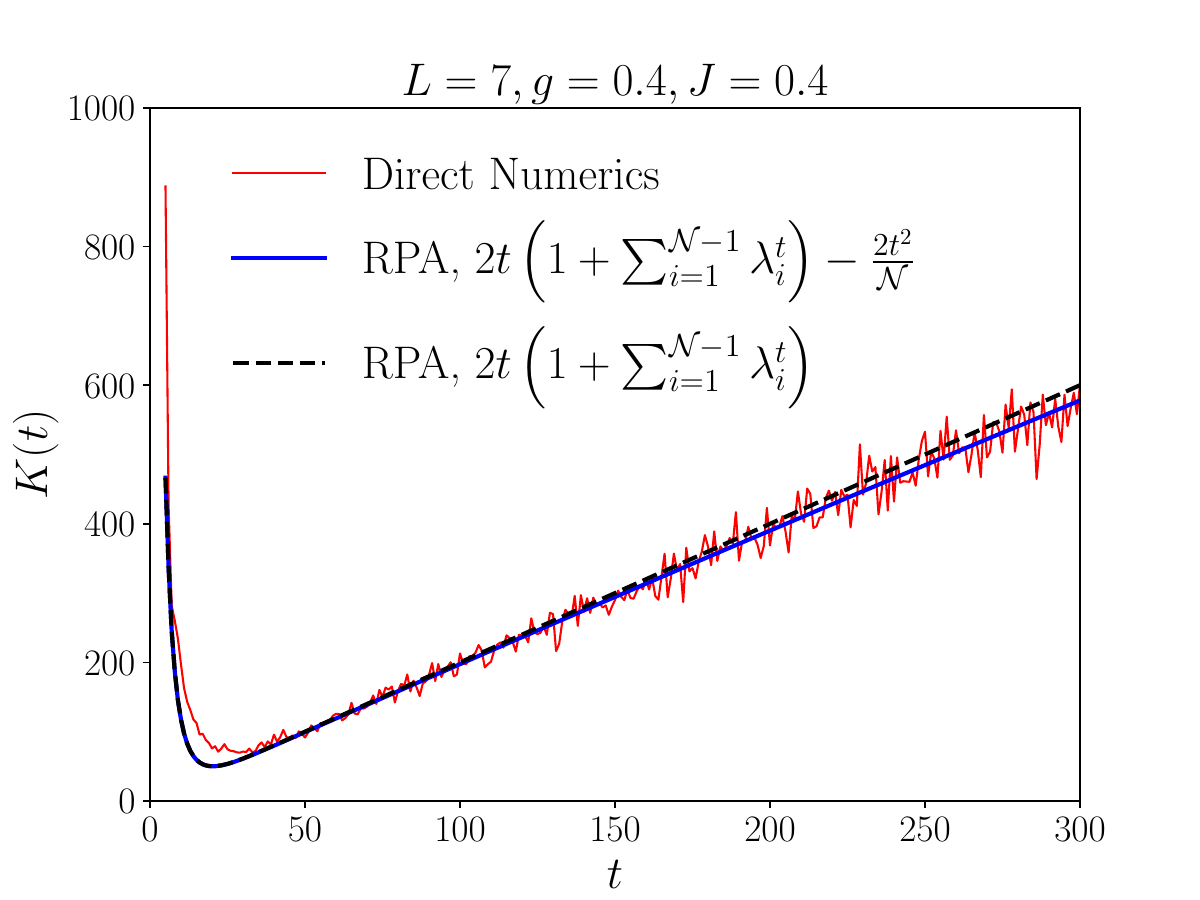}
\caption{Comparison between the exact numerically computed SFF, $K(t)$ vs. $t$, with that obtained using the RPA for Rabi mixing between fermions and qubits. The red curve is exact SFF computed numerically using Eq.~\ref{S4} and the blue (black dashed) curve is that calculated using the first-order and the second-order term (only the first-order term) in time within the RPA. All, $\mathcal{N}_{\text{R}}^{\text F}=8192$, eigenvalues of $\mathcal{M}_{\text{R}}^{\text F}$ are used for the RPA result. For exact numerical computation, we fix $U_0=10,\alpha=1.4$, and $\omega_i,\Omega_i$ are chosen as Gaussian random variables with a mean $\langle\omega_i\rangle=\langle\Omega_i\rangle=1$ and a standard deviation $\sigma_{\omega_i}=\sigma_{\Omega_i}=0.3$. Averaging over $500$ realizations of disorder is performed for the direct SFF computation.}
\label{F5}
\end{figure}

We next numerically compute the SFF for $R$ mixing between fermions and qubits using Eq.~\ref{S4} to compare it to that obtained within the RPA using the full $\mathcal{M}_{\text{R}}^{\text F}$ instead of its Trotter-regime form in Eq.~\ref{S45}. For $R$ mixing violating the total excitation number conservation, we show a better match between two different computations in Fig.~\ref{F5} in comparison to that in Fig.~\ref{F2} for the $JC$ mixing satisfying the total excitation number conservation.

\section{Rabi mixing between bosons and qubits}
\begin{figure}
\centering
\includegraphics[width=\linewidth]{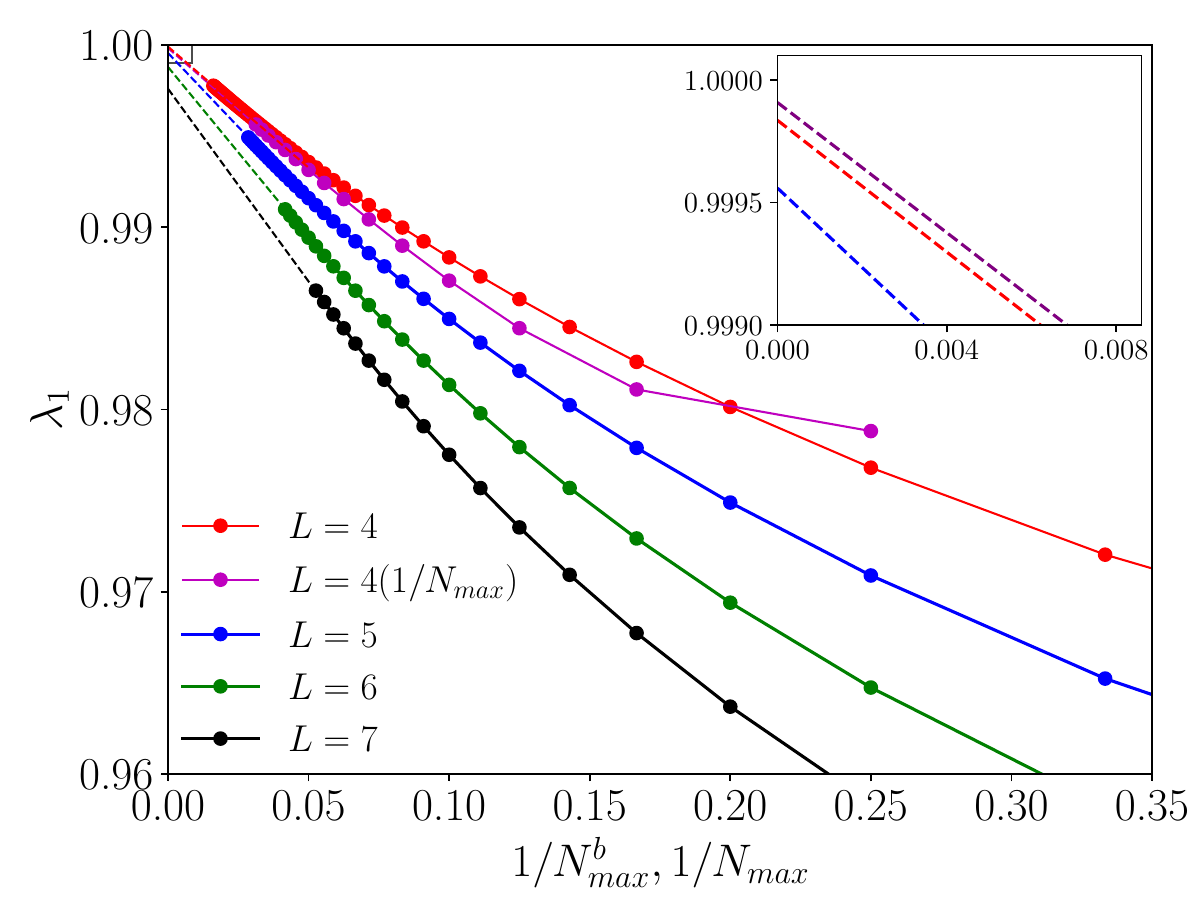}
\caption{Second-largest eigenvalue $\lambda_1$ of $\mathcal{M}_{\text R}^{\text B}$ with inverse maximum number of bosons $(1/N_{max}^b)$ (also inverse maximum number of total excitations $1/N_{max}$ for $L=4$) for four different lengths $(L)$. The dashed curves indicate a linear extrapolation of the last few large $N_{max}^b$ (also $N_{max}$ for $L=4$) points. The parameters are $g=0.1,J=0.4$. The inset  shows a finite gap for $\lambda_1$ as $1/N_{max}^b \to 0$.} 
\label{F6}
\end{figure}

\begin{figure}[H]
\centering
\includegraphics[width=\linewidth]{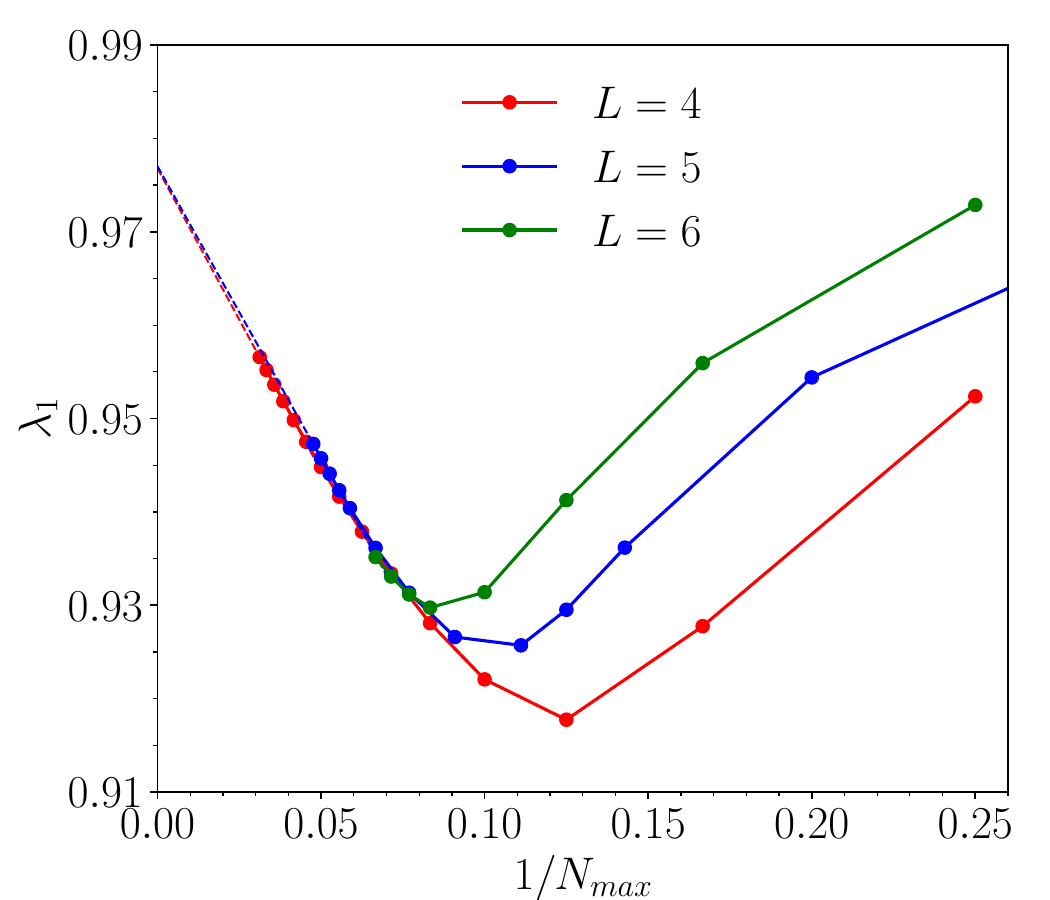}
\caption{Second-largest eigenvalue $\lambda_1$ of $\mathcal{M}_{\text R}^{\text B}$ with inverse maximum number of total excitations $(1/N_{max})$ for three different lengths $(L)$. The dashed curves indicate a linear extrapolation of the last few large $N_{max}$ points. The parameters are $g=0.4,J=0.1$.}
\label{F7}
\end{figure}

The driving Hamiltonian for $R$ mixing between bosons and qubits is the same as $\hat{H}_{\text{R}}$ in Eq.~\ref{S43},  where $\hat{a}_i^\dagger,\hat{a}_i$ are now bosonic creation and annihilation operators. Following the steps presented for deriving the generating Hamiltonian in the Trotter regime for $JC$ mixing between bosons and qubits in Sec.~\ref{JCBoson}, we  get 
\begin{align}
\label{S70}
\mathcal{M}_{\text R}^{\text B}&=\left(1+\frac{J^2L}{2}\right)\mathds{1}_{\mathcal{N}_{\text R}^{\text B}}+2g^2\sum_{i=1}^L\left(\hat{K}_i^1 \hat{\sigma}_i^x+\hat{K}_i^0\right)\notag\\
			&\qquad+2J^2\sum_{i=1}^L\left(\hat{K}_i^1 \hat{K}_{i+1}^1+\hat{K}_i^2 \hat{K}_{i+1}^2-\hat{K}_i^0 \hat{K}_{i+1}^0\right)\notag\\
			&\qquad+\mathcal{O}(J^4,g^4).
\end{align}
The dimension of the Hilbert space $\mathcal{N}_{\text R}^{\text B}$ is infinite for $R$ mixing between bosons and qubits due to the non-conservation of total excitation $N$ and no bound to number of bosons at any site. We could not find the spectrum of $\mathcal{N}_{\text R}^{\text B}$ analytically. Rather, we introduce a truncation to the maximum number of total excitations $N_{max}$ for our numerical study of $\lambda_i$ of $\mathcal{M}_{\text R}^{\text B}$ as in \textcite{RoyPRE2022}. For a fixed $L$, we vary $N_{max}$ to find the asymptotic behavior $(N_{max} \to \infty)$ of the second-largest eigenvalue of $\mathcal{M}_{\text R}^{\text B}$. Following $R$ mixing between fermions and qubits, we explore two different parameter regimes, $g/J\ll 1$ and $g/J\gg1$ in our numerics as given in Figs.~\ref{F6},\ref{F7}.

We further notice that $\mathcal{M}_{\text R}^{\text B}$ commutes with $\hat{\sigma}_i^x$ for all $i\in [1,L]$. Thus, we can label the eigenvalues and eigenvectors of $\mathcal{M}_{\text R}^{\text B}$ by $m_i=(\sigma_i^x+1)/2$, where $\sigma_i^x$ is an eigenvalue of $\hat{\sigma}_i^x$. Nevertheless, we again need to introduce a truncation to the maximum number of bosons $N^b_{max}$ for our numerical study of $\lambda_i$ of $\mathcal{M}_{\text R}^{\text B}$ by fixing the qubit excitations. We find from our numerics for $g=0.1,J=0.4$ that the eigenstates with largest eigenvalue of $\mathcal{M}_{\text R}^{\text B}$ have $m_i=0$ or $m_i=1$ for all $i\in [1,L]$. The degeneracy of $2$ in the largest eigenvalue is due to a $\mathds{Z}_2$ symmetry of $\mathcal{M}_{\text R}^{\text B}$ as like of $\mathcal{M}_{\text R}^{\text F}$. Let us choose the first configuration ($m_i=0$) with the largest eigenvalue. The state corresponding to the second-largest eigenvalue $\lambda_1$ of $\mathcal{M}_{\text R}^{\text B}$ also has the same qubit configuration. The states corresponding to the third-largest eigenvalue are $L$-fold degenerate and nearly degenerate with the second-largest eigenvalue state. The qubit configuration for these states with the third-largest eigenvalue is $m_j=1, m_{i\neq j}=0$ for $i=1,...,j-1,j+1..., L$. This is similar to the results obtained for $R$ mixing between fermions and qubits.

Interestingly, we find from our numerics that the trend of $\lambda_1$ with increasing $N^b_{max}$ matches nicely to that with increasing $N_{max}$ at large values of $N^b_{max}$ and $N_{max}$ when $g/J \ll 1$ as shown in Fig.~\ref{F6} for $L=4$. Since larger $N^b_{max}$ is accessible with a fixed qubit configuration in numerics with a fixed $L$, we employ the numerics by changing $N^b_{max}$ to find  $\lambda_1$ for different $L$'s as $N^b_{max} \to \infty$. We show them in Fig.~\ref{F6} for $L=4,5,6,7$, which depict $\lambda_1$ at asymptotic $N^b_{max}$ slowly drifting towards smaller values with increasing $L$. The decrease in $\lambda_1$ with an increasing $L$ at $N^b_{max} \to \infty$ is probably appearing from the linear extrapolation of the last few large $N^b_{max}$ points, which are not so large for longer $L$. Nevertheless, since the $L$-dependence of $\lambda_1$ is small and there are nearly $(L+1)$-fold degeneracy around the second-largest eigenvalue, we predict $t^*(L) \approx \log L$ when $g/J \ll 1$.  

The similarity between the asymptotic trend of $\lambda_1$ with increasing $N_{max}$ and $N_{max}^b$ is lost when $g/J\gg 1$. Therefore, we can only rely on numerics of $\lambda_1$ with an increasing $N_{max}$ to understand the Thouless-time scaling. We show the behavior $\lambda_1$ with an increasing $N_{max}$ for three different lengths in Fig.~\ref{F7}, which seems to suggest a $L$-independent gap between the first- and second-largest eigenvalues of $\mathcal{M}_{\text R}^{\text B}$ for $g/J\gg 1$. We also observe an $L$-fold degeneracy for the second-largest eigenvalues in our numerics. Therefore, we expect $t^*(L) \approx \log L$ in this regime of parameters too. Nevertheless, our predictions here for the Thouless-time scaling for $R$ mixing between bosons and qubits are based on finite-size numerics with limited data, and there is good scope to improve the current study with more data for more extended system sizes. 

\end{document}